\documentclass[twocolumn,epjc3]{svjour3mod}  

\usepackage[english]{babel}
\usepackage{graphicx}
\usepackage{mathtools, cuted}
\usepackage{amsmath,amssymb,slashed}
\usepackage{multirow}
\usepackage{dsfont}
\usepackage{epstopdf}  
\usepackage[utf8]{inputenc} 
\usepackage[colorlinks,citecolor=blue,urlcolor=blue,linkcolor=blue]{hyperref}

\begin{document}
\title{Electromagnetic transition form factors \\ and Dalitz decays of hyperons}
\author{Nora Salone\thanksref{addr1,addr2} \and Stefan Leupold\thanksref{addr1}}

\institute{Institutionen f\"or fysik och astronomi, Uppsala universitet, Box 516, S-75120 Uppsala, Sweden\label{addr1}
  \and
National Centre for Nuclear Research, 02-093 Warsaw, Poland \label{addr2}}
\date{\today}

\maketitle

\begin{abstract}
    Dalitz decays of a hyperon resonance to a ground-state hyperon and an electron-positron pair can give access to some
    information about the composite structure of hyperons. We present expressions for the multi-differential decay rates in terms of 
    general transition form factors for spin-parity combinations $J^P = \frac12^\pm, \frac32^\pm$ of the hyperon resonance. 
    Even if the spin of the initial hyperon resonance is not measured,
    the self-analyzing weak decay of the ``final'' ground-state hyperon contains information about the relative phase between  
    combinations of transition form factors. 
    This relative phase is non-vanishing because of the unstable nature of the hyperon resonance.
    If all form factor combinations in the differential decay formulae are replaced by their respective values at the photon
    point, one obtains a QED type approximation, which might be interpreted as characterizing hypothetical hyperons with 
    point-like structure. 
    We compare the QED type approximation to a more realistic form factor scenario for the 
    lowest-lying singly-strange hyperon resonances. In this way we explore which accuracy in the measurements of the 
    differential Dalitz decay rates is required
    in order to distinguish the composite-structure case from the pointlike case. Based on the QED type approximation we 
    obtain as a by-product a rough prediction for the ratio between the Dalitz decay width and the corresponding 
    photon decay width.
\end{abstract}

\section{Motivation}
\label{sec:motivation}

Electromagnetic form factors have become an important tool to study the structure of strongly interacting objects, see e.g.\   
\cite{Granados:2017cib,Alarcon:2017asr,Leupold:2017ngs,Junker:2019vvy,Husek:2019wmt,Landsberg:1986fd,Czerwinski:2012ry,%
Miller:2007uy,Pacetti:2015iqa,Punjabi:2015bba,Devenish:1975jd,Korner:1976hv,Carlson:1985mm,Pascalutsa:2006up,Aznauryan:2011qj,Tiator:2011pw,%
Eichmann:2018ytt,Kaxiras:1985zv,Kubis:2000aa,Sanchis-Alepuz:2017mir,Ablikim:2019vaj,Ramalho:2019koj,Ramalho:2020tnn} 
and references therein. 
In the near future, photon and Dalitz decays of hyperons will be measured at GSI/FAIR by HADES+PANDA \cite{Ramstein:2019kaz}, 
$Y^* \to Y \gamma$ and $Y^* \to Y \, e^+ e^-$, 
respectively. Here $Y^*$ denotes a singly-strange hyperon resonance and $Y$ a ground-state hyperon ($\Lambda$ or $\Sigma$).

In the present work we want to explore what it takes to extract more information from the Dalitz decays than from the photon
decays. In other words: How accurately does one need to measure the Dalitz decay distribution to determine an energy dependence
of the form factors? 
In turn, the shape of a form factor is related to the information about the intrinsic structure \cite{Alarcon:2017asr,Miller:2007uy,Tiator:2011pw}. 
To this end, we introduce the most general transition form factors for spin-parity 
combinations $J^P = \frac12^\pm, \frac32^\pm$ of $Y^*$. This is similar in spirit to the developments of 
\cite{Korner:1976hv,Perotti:2018wxm} for $e^+ e^- \to Y^* \bar Y$, but for a different kinematical regime. 
For general considerations about transition form factors 
see also \cite{Devenish:1975jd,Carlson:1985mm}. 
In practice, what we focus on are the low-lying hyperon resonances 
$\Lambda(1405)$ with $J^P=\frac12^-$, $\Lambda(1520)$ with $J^P = \frac32^-$ \cite{pdg}. 
For completeness we cover also the cases of $J^P = \frac32^+$ and $J^P=\frac12^+$. Examples for the latter are the states
$\Sigma(1385)$ and $\Sigma^0$, respectively. Transitions of those states have been studied by the group
of one of the authors in \cite{Granados:2017cib,Junker:2019vvy,Husek:2019wmt,Nair:2018mwa,Holmberg:2018dtv}. 

The $\Lambda(1405)$ is a very interesting state; see e.g.\ \cite{Mai:2020ltx,Dalitz:1967fp,Siegel:1988rq,Jido:2003cb,GarciaRecio:2003ks,Geng:2007hz,Sekihara:2008qk,Hall:2014uca} and references therein. 
It is the lowest-lying baryon state with negative parity \cite{pdg}. \newline Naively, one would
think that the lowest-lying baryon with negative parity should contain only up and down quarks since those are significantly
lighter than the \textit{s}-quark. Yet, the $\Lambda(1405)$ as the lightest baryon with negative parity has a stangeness
of $-1$, i.e.\ must contain (at least) one strange quark.
This triggered many discussions about the nature of the $\Lambda(1405)$ as a state that might not fit into the
quark model, which describes baryons as three-quark states. As an alternative, a bound state of nucleon and antikaon
(``hadronic molecule'') has been
proposed. It has also been suggested that there might actually be two coupled-channel hadronic-molecule states \cite{Jido:2003cb}, one coupling
stronger to the nucleon-antikaon structure and one stronger to $\Sigma$-pion, which constitutes the main decay mode of the $\Lambda(1405)$. 
From a more quantitative point of view the proper question is how much overlap a physical $\Lambda(1405)$ state
has e.g.\ with a three-quark or with a proton-antikaon field configuration etc. In any case, it is conceivable that different
pictures about the nature of the $\Lambda(1405)$ lead to somewhat different size predictions. More generally, different ideas 
about the intrinsic structure can lead to different predictions for the differential Dalitz decay rates.

The second concrete example for which we will provide quantitative results is the $\Lambda(1520)$. It is the strange baryon
next in mass after covering $\Sigma$ and $\Sigma(1385)$ 
previously \cite{Granados:2017cib,Junker:2019vvy,Husek:2019wmt,Nair:2018mwa,Holmberg:2018dtv} and the $\Lambda(1405)$ in the
present work. Above the $\Lambda(1520)$ our low-energy techniques might fail to work. To provide a self-contained work, 
it makes sense to cover also the $\Lambda(1520)$ in the present paper. 

Though the $\Lambda(1520)$ is often
regarded as a typical quark-model state, see e.g.\ the mini-review in \cite{pdg}, there are also ideas that
suggest the $\Lambda(1520)$ essentially as a hadron molecule partner to the $\Lambda(1405)$ when interchanging
ground-state baryons from the nucleon octet by spin-3/2 states from the $\Delta$ decuplet \cite{Kolomeitsev:2003kt}.
Like for the $\Lambda(1405)$, it can
be expected that different pictures about the structure of the $\Lambda(1520)$ \cite{Roca:2006pu,Roca:2006sz}
lead to different predictions for the differential Dalitz decay rates. 

The purpose of the present paper is not about developing particular models for the structure of the $\Lambda(1405)$ or the
$\Lambda(1520)$. Yet we take the large interest in these states as a motivation to perform a model-independent analysis of the
capability of Dalitz decays to access their respective intrinsic structure.

The paper is structured in the following way. In the next section we present the framework to introduce 
electromagnetic transition form factors in the most general way for initial states with spin 3/2 or 1/2 and final states with
spin 1/2. We will calculate the decay rates for radiative (photon) decays and Dalitz decays and also include the possibility that
the ``final'' hyperon emerging from the Dalitz decay performs a further weak decay into nucleon and pion.
In sect.\ \ref{sec:FFparam} we introduce our parametrization for the transition form factors that features transition radii.
We also specify a ``structureless'' case that we can contrast with the case of an extended structure.
In sect.\ \ref{sec:concrete} we apply our framework concretely to the two initial
states $\Lambda(1405)$ and $\Lambda(1520)$ (and the final state of a ground state $\Lambda$). Further discussion and a
summary is provided in sect.\ \ref{sec:summary}. Appendices are added for technical purposes, but also to explain some conceptual
issues that do not fit into the main body of the text.

\section{Constraint-free form factors, helicity amplitudes and differential decay widths}
\label{sec:calcs}

From a formal point of view we study electromagnetic transitions from a baryon resonance with spin 1/2 or 3/2 to a 
baryon with spin 1/2. 
For this transition, we disregard parity violating processes, i.e.\ we focus on transitions mediated by the strong and 
electromagnetic interactions. 
Prominent examples that can be described by this framework are the decays:\newline
$\Lambda(1520) \to \Lambda \gamma^{(*)}, \Sigma^0 \gamma^{(*)}$; 
$\Lambda(1405) \to \Lambda \gamma^{(*)}, \Sigma^0 \gamma^{(*)}$; $\Sigma(1385) \to \Lambda \gamma^{(*)}, \Sigma \gamma^{(*)}$; 
$\Sigma^0 \to \Lambda \gamma^{(*)}$. \newline Electroweak processes like $\Xi^0 \to \Lambda \gamma^{(*)}$ would need a (straightforward) 
extension of the formalism and are not covered in the present work. We will present formulae for all parity combinations. When 
it comes to concrete applications we will focus on two processes, namely $\Lambda(1520) \to \Lambda \gamma^{(*)}$ and 
$\Lambda(1405) \to \Lambda \gamma^{(*)}$.  

Generically we study the process $Y^* \to Y \gamma^{(*)}$ where the star for the hyperon $Y$ denotes an excited hyperon, i.e.\ 
a resonance, while $\gamma^*$ refers to a virtual photon $\gamma$.

\subsection{Transition $\frac{3}{2}^{\mp} \to \frac{1}{2}^{\pm}$} 
\label{sec:opp-par3/2}
If the initial baryon $Y^*$ has spin 3/2 and opposite parity to the final baryon $Y$, the most general decomposition of the 
transition respecting Lorentz invariance, current conservation and parity symmetry can be written as 
(cf.\ also \cite{Korner:1976hv})
    \begin{equation}
    \resizebox{0.92\hsize}{!}{%
        $\langle p_Y,\lambda_Y \vert j^\mu(0) \vert p_{Y^*},\lambda_{Y^*} \rangle = 
    e \bar{u}(p_{Y},\lambda_{Y}) \Gamma^{\mu\nu}_- u_\nu(p_{Y^*},\lambda_{Y^*})
    \label{eq:3/2-}$%
        }
\end{equation}
with 
\begin{equation}
    \begin{split}
        \qquad \Gamma^{\mu\nu}_- =& -i H_1(q^2) \, m_{Y^*} \, \left(\gamma^{\mu } q^{\nu }-\slashed{q} \, g^{\mu  \nu } \right)  \\
    & +i H_2(q^2) \left(q^{\nu} p_{Y^*}^{\mu}-(q \cdot p_{Y^*}) \, g^{\mu  \nu } \right)  \\
    & +i H_3(q^2) \left(q^{\mu} q^{\nu }-q^2 g^{\mu  \nu }\right)
    \end{split}
    \label{eq:3/2-Gam}
\end{equation}
and $q:= p_{Y^*} - p_Y$.
Here $j_\mu$ denotes the electromagnetic current and $e$ the charge of the proton. The helicity of the initial (final) baryon is 
denoted by $\lambda_{Y^*}$ ($\lambda_{Y}$). Our conventions for the spin-3/2 vector-spinor 
$u_\nu$ have been spelled out in \cite{Junker:2019vvy}. 
Note that no $\gamma_5$ appears here since either none or both of the involved baryons have natural 
parity \cite{Devenish:1975jd} (and the electromagnetic current is a vector current and has natural parity). 
The three quantities $H_i$, $i=1,2,3$, constitute constraint-free transition form factors in the sense of a Bardeen-Tung-Tarrach 
(BTT) construction \cite{Bardeen:1969aw,Tarrach:1975tu}. 

We have introduced the three transition form factors $H_i$ such that they all have the same 
dimensionality (two inverse mass dimensions). 
In general, these transition form factors are complex quantities. Thus the appearance of the explicit $i$'s in the defining 
eq.\ \eqref{eq:3/2-Gam} is a pure convention. However, there is some meaning to this choice. Suppose one calculates 
contributions to the transition form factors from an effective Lagrangian that satisfies charge conjugation symmetry. Then, a 
tree-level calculation will yield purely real results for $H_i$. In other words, conventions have been chosen such that only 
loops create imaginary parts for $H_i$. We substantiate this further in \ref{sec:effL}.

Next, we introduce dimensionless helicity amplitudes: we define 
\begin{eqnarray}
    H_-(q^2) &:=& - \left(m_{Y^*}-m_{Y}\right) m_{Y^*}  H_1(q^2)  \nonumber \\
    && {} +\frac{1}{2}  \left(m_{Y^*}^2-m_{Y}^2+q^2\right) H_2(q^2)+ q^2 H_3(q^2) \,, \nonumber \\
    H_0(q^2) &:=& - \ (m_{Y^*}-m_{Y}) \, m_{Y^*} H_1(q^2)  \nonumber \\
    \label{eq:defhelampl32-}
    && {} +(m_{Y^*}-m_{Y}) \, m_{Y^*} H_2(q^2)  \\
    && {} + \frac{m_{Y^*}-m_{Y}}{2 m_{Y^*}} \left(m_{Y^*}^2-m_{Y}^2+q^2\right) H_3(q^2) \,, \nonumber \\
    H_+(q^2) &:=& -\left(m_{Y^*} m_{Y}-m_{Y}^2+q^2 \right) H_1(q^2)  \nonumber \\
    && {} +\frac{1}{2}  \left(m_{Y^*}^2-m_{Y}^2+q^2\right) H_2(q^2) + q^2 H_3(q^2) \,. \nonumber 
\end{eqnarray}
In a frame where the baryon momenta are aligned, the helicity flip amplitude $H_-$ is related to the combinations 
$(\lambda_{Y^*},\lambda_Y) = (3/2,1/2), (-3/2,-1/2)$. 
The other helicity flip amplitude $H_+$ is related to the combinations 
$(\lambda_{Y^*},\lambda_Y) = (-1/2,1/2), (1/2,-1/2)$. Finally the non-flip amplitude $H_0$ relates to 
$\lambda_{Y^*}=\lambda_{Y}=\pm 1/2$. Note that the spin\footnote{We avoid here the phrase ``helicity'' since the virtual photon
might be at rest.} of the real or virtual photon along the axis defined by the
flight direction of the hyperons is given by $\lambda_\gamma=\lambda_{Y^*}-\lambda_Y$. 

The advantage of the helicity amplitudes over the three transition form factors $H_i$, $i=1,2,3$, is that
there are no interference terms when calculating the decay widths for the reactions $Y^* \to Y \gamma$ and
$Y^* \to Y \, e^+ e^-$. We will see this explicitly below. 
A second advantage is that one can use quark counting rules to determine the high-energy behavior \cite{Carlson:1985mm}
of the helicity amplitudes for large values of space-like $q^2 <0$. The main topic of this work are Dalitz decays. Here the 
photon virtuality $q^2$ is time-like and has an upper limit given by $(m_{Y^*}-m_Y)^2$. Therefore high-energy constraints are 
not so relevant for the physics discussed here. Nonetheless, for completeness, we collect the high-energy behavior of all 
transition form factors and helicity amplitudes in \ref{sec:high-qcr}. 

While the transition form factors $H_1$, $H_2$ and $H_3$ are free from kinematical constraints, the helicity amplitudes 
satisfy
\begin{equation}
    \begin{split}
        \qquad H_+((m_{Y^*}-m_{Y})^2) &= H_0((m_{Y^*}-m_{Y})^2) \\ 
        &= H_-((m_{Y^*}-m_{Y})^2)
    \end{split}
  \label{eq:helconstm3/2}
\end{equation}
and
\begin{equation}
    \begin{split}
        \qquad &\frac{2(m_{Y^*}+m_{Y})}{m_{Y^*}-m_{Y}} H_0((m_{Y^*}+m_{Y})^2)  \\ 
        &= H_+((m_{Y^*}+m_{Y})^2) + H_-((m_{Y^*}+m_{Y})^2)  \,. 
    \end{split}
  \label{eq:helconstm3/2-high}
\end{equation}
From a technical point of view, these relations are easy to deduce from the definitions of the 
helicity amplitudes \eqref{eq:defhelampl32-}. Later, the constraint \eqref{eq:helconstm3/2} will be very important for our model independent low-energy parametrization of the transitions. Therefore we feel obliged to offer also a physical instead of purely technical motivation for the kinematical constraints. This physical explanation is provided in \ref{sec:multi-p}. 

The width for the two-body radiative decay \newline $Y^* \to Y \gamma$ is given by 
\begin{multline}
    \Gamma_2 = e^2 \frac{\left(m_{Y^*}+m_{Y}\right)^2 \left(m_{Y^*}^2-m_{Y}^2\right)}{96 \pi  m_{Y^*}^3} \\ 
   \times \left[3| H_-(0)|^2 + | H_+(0)|^2 \right]%
   \,.
  \label{eq:radiative32-}
\end{multline}
As already announced there are no interference terms between the helicity amplitudes. The absence of $H_0$ signals nothing but the non-existence of a longitudinally polarized real photon. 

To describe in the one-photon approximation the Dalitz decay $Y^* \to Y \gamma^* \to Y \, e^+ e^-$ we choose a frame where the virtual photon (and therefore the electron-positron pair) is at rest. In this frame, $\theta$ denotes the angle between the hyperon $Y$ and the electron. The  double-differential three-body (Dalitz) decay width is given by
\begin{equation} 
\scalebox{0.96}{%
$\begin{split}
        &\frac{\text{d}\Gamma_3}{\text{d}q^2\text{d}(\cos{\theta})} = \frac{e^4 p_z \sqrt{q^2} \beta_e}{(2\pi)^3  \, 192 m_{Y^*}^3}\frac{(m_{Y^*}+m_{Y})^2-q^2}{q^2}   \\
  & \hspace{1mm} \times \bigg\{\bigg(1+ \cos ^2\theta+ \frac{4 m_e^2}{q^2} \sin ^2\theta\bigg) \big[3 |H_-(q^2)|^2 + |H_+(q^2)|^2 \big] \\
  & \hspace{2mm} +\bigg(\sin ^2\theta+ \frac{4 m_e^2}{q^2}\cos^2\theta\bigg) \frac{4q^2}{(m_{Y^*}-m_{Y})^2} |H_0(q^2)|^2\bigg\} \, .
    \end{split}$}
    \label{eq:dalitz32-}
\end{equation}

Here we have used the velocity of the electron in the rest frame of the electron-positron pair given by 
$\beta_e = \sqrt{1-4m_e^2/q^2}$. The momentum of $Y^*$ and $Y$ in the rest frame of the virtual photon is given by
\begin{equation}
  \label{eq:pzdef}
 \qquad \qquad \qquad p_z := \frac{\lambda^{1/2}(m_{Y^*}^2,m_Y^2,q^2)}{2 \sqrt{q^2}} 
\end{equation}
with the K\"all\'en function
\begin{equation}
  \label{eq:kallenfunc}
  \qquad\lambda(a,b,c):=a^2+b^2+c^2-2(ab+bc+ac) \,.
\end{equation}

To obtain the integrated Dalitz decay width, we note that the lower (upper) integration limit of the $\cos\theta$ integration is 
$-1$ ($+1$), i.e.\ the $\theta$ integration would go from $\pi$ to 0 and not the other way. This convention makes the 
right-hand side of \eqref{eq:dalitz32-} positive. 

It is nice to see how all physical constraints are visible in \eqref{eq:dalitz32-}. For $q^2 =(2 m_e)^2$ the leptons are produced at rest.
Thus one cannot define an angle between electron and hyperon momentum. The right-hand side of \eqref{eq:dalitz32-} should
show no angular dependence. This is indeed the case. At the other end of the phase space, i.e.\ for $q^2 =(m_{Y^*}-m_Y)^2$
the hyperons are at rest. Again one cannot define an angle between electron and hyperon momentum. Using the kinematical
constraint \eqref{eq:helconstm3/2}, one can see that also here the angular dependence disappears. 

Even though the electromagnetic transition that we consider respects parity symmetry, we can allow for a further decay of the 
``final'' hyperon. Focusing now on $Y=\Lambda$ (or $\Sigma^\pm$) this last decay is mediated by the weak interaction and does 
violate the parity symmetry. As a consequence, this decay populates different partial waves and gives rise to an interference 
pattern. In total, we study now the decay sequence $Y^* \to Y \, e^+ e^-$ and $Y \to \pi N$. Thus we have a four-body 
final state. In principle, this gives rise to 5 independent kinematical variables. However, the intermediate $Y$ state is 
so long-living \cite{pdg} that its mass is fixed (and in the experimental analyses one triggers 
on a displaced vertex \cite{Thome:2012bdy,IkegamiAndersson:2020rau}). 
In addition, one can show that the squared matrix element of the four-body decay is independent of specific combinations of 
four-momenta. This feature is discussed in \ref{app:4body}. As a consequence, the specific four-body decay depends 
on three independent variables, one variable more than the already considered Dalitz decay. 

Besides the invariant mass $\sqrt{q^2}$ of the dilepton pair (the photon virtuality $\gamma^*$) and the angle $\theta$ between the electron and the hyperons in the rest frame 
of $\gamma^*$, one could use a second relative angle. 
It is convenient to define this angle in the frame where the decaying $Y$ hyperon is at rest. In this frame the electron-positron
pair defines a plane. One could use the angle between the plane's normal and the direction of the nucleon. Since it is a relative 
angle, its definition does not depend on the choice of a coordinate system, only on the choice of a proper frame of reference. 
Yet, to connect to the formalism 
developed in \cite{Perotti:2018wxm,Faldt:2013gka,Faldt:2016qee,Faldt:2017kgy,Faldt:2017yqt,Faldt:2019zdl} we introduce a 
fixed coordinate system and find for 
the four-body decay $Y^* \to Y \gamma^* \to \pi N \, e^+ e^-$ the differential decay width 
\begin{equation}
\scalebox{0.96}{%
$\begin{split}
        &\frac{\text{d}\Gamma_4}{\text{d}q^2\text{d}(\cos{\theta})\text{d}\Omega_N} \\
        &= \frac{e^4 p_z \sqrt{q^2} \beta_e }{(2\pi)^4 \, 384 m_{Y^*}^3} \, \text{Br}_{Y \to \pi N}
  \, \frac{(m_{Y^*}+m_{Y})^2-q^2}{q^2}      \\
  & \hspace{1mm} \times \bigg\{\bigg(1+ \cos ^2\theta+ \frac{4 m_e^2}{q^2} \sin ^2\theta\bigg)\big[|H_+(q^2)|^2 + 3 |H_-(q^2)|^2 \big]   \\
  &\hspace{5mm} +\bigg(\sin^2\theta+ \frac{4 m_e^2}{q^2}\cos^2\theta\bigg) 
  \frac{4q^2}{\left(m_{Y^*}-m_{Y}\right)^2} |H_0(q^2)|^2    \\
  &\hspace{5mm} -\frac{4\sqrt{q^2} \beta_e^2 }{m_{Y^*}-m_{Y}} \, \alpha _{Y} \, \text{Im}[H_0(q^2) H_+^*(q^2)]   \\
  &\hspace{9mm}\times \sin \theta \cos \theta \sin \theta_N \sin \phi_N \bigg\}   \,.
    \end{split}$}
    \label{eq:fourbody-32-}
    \raisetag{15pt}
\end{equation}
We recall that $q^2$ is the square of the dilepton mass or photon virtuality 
and $\theta$ denotes the angle between one of the hyperons and the electron in the rest frame of the dilepton (rest frame of the
virtual photon). 
In addition, $\theta_N$ and $\phi_N$ are the angles of the nucleon three-momentum measured in the rest frame of $Y$.
The coordinate system in this frame is defined by $\vec q$ pointing in the negative $z$-direction
(i.e.\ in the rest frame of the virtual photon the $Y$ direction defined the positive $z$-axis) and the electron moves
in the $x$-$z$ plane with positive momentum projection on the $x$-axis. In this frame, $\theta_N$ is the angle of the nucleon  
momentum relative to the $z$-axis and $\phi_N$ is the angle between the $x$-axis and the projection of the nucleon momentum
on the $x$-$y$ plane, i.e.
\begin{align}
     \qquad  \textbf{p}_N = p_{\rm f} \, (\sin\theta_N &\cos\phi_N,\sin\theta_N \sin\phi_N,\cos\theta_N) \,, \nonumber \\
    \hspace{2cm} \textbf{q} &= \vert \textbf{q} \vert \, (0,0,-1)  \,, \label{eq:defmoms-Lambdarest} \\
     \textbf{p}_{e^-} \cdot \textbf{e}_y = 0 \,, \quad & \textbf{p}_{e^-} \cdot \textbf{e}_x > 0 \,, \quad
    \textbf{e}_y = \frac{\textbf{p}_{e^-} \times \textbf{q}}{\vert \textbf{p}_{e^-} \times \textbf{q} \, \vert} \nonumber \,,
\end{align}
with the momentum $p_{\rm f}$ of the nucleon in the rest frame of the decaying $Y$ hyperon. We provide the momentum and 
also the corresponding energy:
\begin{equation}
  \label{eq:protonen2}
  \qquad \qquad \qquad p_{\rm f} = \frac{\lambda^{1/2}(m_Y^2,m_N^2,m_\pi^2)}{2 m_Y}
\end{equation}
and 
\begin{equation}
  \label{eq:protonen}
  \qquad \qquad \qquad E_N = \frac{m_Y^2+m_N^2-m_\pi^2}{2 m_Y}   \,,
\end{equation}
with the K\"all\'en function defined in \eqref{eq:kallenfunc}. 
Note the subtlety that $\theta$ is measured in the rest frame of the virtual photon while $\Omega_N$ denotes angles in the
rest frame of the $Y$ hyperon. In terms of Lorentz invariant quantities the angles are related to
\begin{eqnarray}
  p_Y \cdot k_e &=& -\frac12 \, \lambda^{1/2}(m_{Y^*}^2,m_Y^2,q^2) \, \beta_e \, \cos\theta \,, \nonumber \\
  \epsilon_{\mu\nu\alpha\beta} \, k_e^\mu \, p_Y^\nu \, p_N^\alpha \, q^\beta
                      &=& - \frac12 \, \sqrt{q^2} \, \lambda^{1/2}(m_{Y^*}^2,m_Y^2,q^2) \, p_{\rm f} \, \beta_e \nonumber \\ 
                      && \times  \sin\theta \sin\theta_N \sin\phi_N
  \label{eq:LIangles}
\end{eqnarray}
with $k_e := p_{e^-}-p_{e^+}$, $q = p_{e^-}+p_{e^+} = p_{Y^*}-p_Y$ and the convention \cite{pesschr}
for the Levi-Civita symbol: $\epsilon_{0123}=-1$. 
Other aspects of the four-body decay are discussed in \ref{app:4body}.

The final weak decay of a spin-1/2 hyperon to a nucleon and a pion is driven by the matrix element \cite{pdg}
\vspace{-10pt}
\begin{equation}
  \label{eq:Lambdadecay}
  \qquad {\cal M}_{\rm weak} = G_F \, m_\pi^2 \, \bar u_N(p_N) \left(A-B \gamma_5 \right) u_Y(p_Y)   \,.
\end{equation}
It is useful to introduce the asymmetry parameter
\begin{equation}
  \qquad\qquad \qquad \alpha_Y  :=  \frac{2 {\rm Re}(T^*_s T_p)}{\vert T_s \vert^2 + \vert T_p \vert^2}
  \label{eq:defasym}
\end{equation}
with the \textit{s}-wave amplitude $T_s := A$, the \textit{p}-wave amplitude $T_p := p_{\rm f} \, B/(E_N+m_N)$ 
and mass $m_N$, energy $E_N$ and 
momentum $p_{\rm f}$ of the nucleon in the rest frame of the decaying hyperon $Y$; see also appendix A of \cite{Faldt:2013gka} 
for further details that are useful for practical calculations. 

For stable particles, e.g.\ nucleons, the electromagnetic form factors are complex for positive transferred 
momentum $q^2\ge 4m_N^2$ i.e.\ for the reaction $e^+e^- \to N\bar{N}$ in the time-like region of $q^2$. On the other hand, the 
form factors are real for the space-like region $q^2<0$, i.e.\ for the scattering process $e^-N\to e^-N$. 
However, for resonances, e.g.\ $Y^*$, the TFFs are complex for all values of $q^2$ \cite{Junker:2019vvy}. 
Therefore the interference terms in \eqref{eq:fourbody-32-} can in principle be measured. 
They contain information that is complementary to the moduli that are accessible by the Dalitz decay parametrized by 
\eqref{eq:dalitz32-}. A calculation of such interference terms is beyond the scope of this work.
Yet, the results of \cite{Junker:2019vvy} suggest that such interference terms are relatively small. 
Note that the spin-parity combination considered in \cite{Junker:2019vvy} refers 
strictly speaking to subsection \ref{sec:same-par3/2} below, but semi-quantitatively we expect a similar pattern. 
High experimental accuracy will be required to access these interference terms in the Dalitz decay region. Yet, it 
might be worth to extract this additional structure information. It would also be interesting to see how different models 
for the structure of hyperon resonances differ in their predictions for such interference terms. In this context we stress once 
more that in practice these interference terms are driven by the fact that resonances are unstable with respect to the strong interaction. Quasi-stable states (which decay only because of the electromagnetic or weak interaction) would have tiny imaginary parts of form factors. For the same reason, models that treat resonances as stable are not capable to provide predictions for such interference terms. For measurements of the latter in the production region of hyperon-antihyperon pairs see \cite{Ablikim:2019vaj}. Note also that one needs an additional (weak) decay (more generally a decay that populates more than one partial wave) to make the interference term visible. If the asymmetry parameter $\alpha_Y$ vanished, one would not see an interference effect in (\ref{eq:fourbody-32-}).

\subsection{Transition $\frac{1}{2}^{\mp} \to \frac{1}{2}^{\pm}$}
\label{sec:opp-par1/2}

The structure of this subsection follows closely the previous one. 
If the initial baryon $Y^*$ has spin 1/2 and opposite parity to the final baryon $Y$, the most general decomposition of the 
transition respecting Lorentz invariance, current conservation and parity symmetry can be written as 
(cf.\ also \cite{Korner:1976hv})
\begin{equation}
    \resizebox{0.9\hsize}{!}{%
        $\langle p_Y,\lambda_Y \vert j^\mu(0) \vert p_{Y^*},\lambda_{Y^*} \rangle = 
    e \bar{u}(p_{Y},\lambda_{Y}) \, \Gamma^{\mu}_- \, u(p_{Y^*},\lambda_{Y^*})\label{eq:1/2-}$%
        }
\end{equation}
with
\begin{equation}
     \resizebox{0.9\hsize}{!}{%
      $\Gamma^\mu_- = \tilde{F}_2(q^2) \, m_{Y^*} \, \sigma^{\mu\beta} q_\beta \gamma_5
    + i\tilde{F}_3(q^2) \left(q^2\gamma^\mu - \slashed{q}q^\mu \right)\gamma_5 \,.
    \label{eq:vertfunc12m}$%
    }
\end{equation}
Note the appearance of a $\gamma_5$ since one of the baryons has unnatural parity. 
The two quantities $\tilde F_2$ and $\tilde F_3$ constitute constraint-free transition form factors in the sense of a 
BTT construction \cite{Bardeen:1969aw,Tarrach:1975tu}. The labeling is motivated in \ref{sec:effL}.

We introduce dimensionless helicity amplitudes:
\begin{align}
    \qquad \tilde{F}_0(q^2) :={} & (m_{Y^*}-m_{Y})^2 \tilde{F}_3(q^2)  \nonumber \\
    & {}- (m_{Y^*}-m_{Y}) \, m_{Y^*} \, \tilde F_2(q^2)  \,,   \label{eq:defhelm1/2}
    \\ 
    \qquad \tilde{F}_+(q^2) :={} & q^2 \tilde{F}_3(q^2)- (m_{Y^*}-m_{Y})\, m_{Y^*} \, \tilde{F}_2(q^2)  \,. \nonumber
\end{align}
In a frame where the baryon momenta are aligned, the helicity flip amplitude $\tilde F_+$ is related to the combinations 
$(\lambda_{Y^*},\lambda_Y) = (-1/2,1/2), (1/2,-1/2)$. The non-flip amplitude $\tilde F_0$ relates to 
$\lambda_{Y^*}=\lambda_{Y}=\pm 1/2$. 

While the transition form factors $\tilde F_2$ and $\tilde F_3$ are free from kinematical constraints, the helicity amplitudes satisfy 
\begin{equation}
  \qquad\tilde{F}_+((m_{Y^*}-m_{Y})^2) = \tilde{F}_0((m_{Y^*}-m_{Y})^2)  \,.
  \label{eq:helconstm1/2}
\end{equation}
We will use this later for a model independent low-energy parametrization of the transitions. 
\\The constraint \eqref{eq:helconstm1/2} can be easily deduced from the definitions \eqref{eq:defhelm1/2}. Physically it 
follows from the fact that one partial wave is dominant over the other at the end of the phase space of the Dalitz decay 
$Y^* \to Y \, e^+ e^-$; see the related discussion in \ref{sec:multi-p}. 

The respective decay widths for the two-body radiative decay $Y^* \to Y \gamma$, the three-body Dalitz decay 
$Y^* \to Y \gamma^* \to Y \, e^+ e^-$,
and the four-body decay $Y^* \to Y \gamma^* \to \pi N \, e^+ e^-$ are given by
\begin{equation}
  \quad\Gamma_2 = \frac{e^2 | \tilde{F}_+(0)|^2 \left(m_{Y^*}+m_{Y}\right)^2 \left(m_{Y^*}^2-m_{Y}^2\right)}{8 \pi  m_{Y^*}^3}  \,,
  \label{eq:real1/2-}
\end{equation}
\begin{align}
    &\frac{\text{d}\Gamma_3}{\text{d}q^2\text{d}(\cos{\theta})} = 
        \frac{e^4 p_z \sqrt{q^2}\beta_e}{(2\pi)^3 \, 16 m_{Y^*}^3}\frac{(m_{Y^*}+m_{Y})^2-q^2}{q^2} \nonumber\\
        & \quad \times \bigg\{\bigg(1+\cos^2{\theta}+\frac{4m_e^2}{q^2}\sin^2{\theta} \bigg) \, |\tilde{F}_+(q^2)|^2 \label{eq:12mddiff} \\
        & \hspace{-2mm}\qquad +\bigg(\sin^2{\theta}+\frac{4m_e^2}{q^2}\cos^2{\theta} \bigg) \frac{q^2}{(m_{Y^*}-m_{Y})^2}|\tilde{F}_0(q^2)|^2\bigg\}  \,, \nonumber
\end{align}
and 
\begin{equation}
    \scalebox{0.98}{%
        $\begin{split}
            & \frac{\text{d}\Gamma_4}{\text{d}q^2\text{d}(\cos{\theta})\text{d}\Omega_N}\\
        &= \frac{e^4 p_z \sqrt{q^2} \beta _e}{(2\pi)^4 \, 32 m_{Y^*}^3} \text{Br}_{Y \to \pi N} \, \frac{(m_{Y^*}+m_{Y})^2-q^2}{q^2}  \\
        & \hspace{2mm} \times \bigg\{\bigg(1+ \cos ^2\theta+ \frac{4 m_e^2}{q^2} \sin ^2\theta\bigg) |\tilde{F}_+(q^2)|^2  \\
        & \hspace{3mm} +\bigg(\sin ^2\theta+ \frac{4 m_e^2}{q^2}\cos ^2\theta\bigg) 
        \frac{q^2}{(m_{Y^*}-m_{Y})^2} |\tilde{F}_0(q^2)|^2  \\
        &\hspace{3mm} +\frac{2\sqrt{q^2} \beta_e^2}{m_{Y^*}-m_{Y}} \, \alpha _{Y} \, 
        \text{Im}[\tilde{F}_0(q^2) \tilde{F}_+^*(q^2)]  \\
        & \qquad \times \sin \theta \cos \theta \sin \theta_N \sin \phi_N  \bigg\}   \,.
        \end{split}$}
  \label{eq:weak12m}
\end{equation}

Again, it is nice to see how all angular dependence disappears for the cases where specific three-vectors vanish and therefore 
do not allow to define relative angles. For $q^2 = (2 m_e)^2$ one can see directly how all angular dependence vanishes in the 
curly brackets of \eqref{eq:12mddiff} and \eqref{eq:weak12m}. For $q^2 = (m_{Y^*}-m_{Y})^2$ this is achieved by the kinematical 
constraint \eqref{eq:helconstm1/2}.

\subsection{Transition $\frac{3}{2}^{\pm} \to \frac{1}{2}^{\pm}$}
\label{sec:same-par3/2}

The previous two subsections were devoted to cases that we will study in more detail later. For completeness, 
we also add two more combinations in the present and the succeeding subsection.

If the initial baryon $Y^*$ has spin 3/2 and the same parity as the final baryon $Y$, the corresponding decomposition is given by
\begin{equation}
    \resizebox{0.91\hsize}{!}{%
    $\langle p_Y,\lambda_Y \vert j^\mu(0) \vert p_{Y^*},\lambda_{Y^*} \rangle = 
    e \bar{u}(p_{Y},\lambda_{Y}) \Gamma^{\mu\nu}_+ u_\nu(p_{Y^*},\lambda_{Y^*})
    \label{eq:3/2+}$}
\end{equation}
\begin{equation}
    \begin{split}
       \qquad \Gamma^{\mu\nu}_+ ={}& \tilde H_1(q^2) \, m_{Y^*} \left(\gamma^{\mu } q^{\nu }-\slashed{q}g^{\mu  \nu } \right) \gamma_5  \\ 
    & + \tilde H_2(q^2) \left(q^{\nu} p_{Y^*}^{\mu} -(q \cdot p_{Y^*})  g^{\mu\nu} \right) \gamma_5  \\ 
    & + \tilde H_3(q^2) \left(q^{\mu} q^{\nu} -q^2 g^{\mu\nu} \right) \gamma_5 \,.
    \end{split}
    \label{eq:3/2+Gamma}
\end{equation}
We see the appearance of a $\gamma_5$ since one of the baryons has unnatural parity. 
The three quantities $\tilde H_i$, $i=1,2,3$ constitute constraint-free transition form factors in the sense of a 
BTT construction \cite{Bardeen:1969aw,Tarrach:1975tu}. 

We note that formally $\Gamma^{\mu\nu}_+$ is obtained 
from \eqref{eq:3/2-Gam} by multiplying with $-i\gamma_5$ from the left \cite{Korner:1976hv} and changing $H \to \tilde H$. 
One can rephrase this by stating that one obtains \eqref{eq:3/2+}, \eqref{eq:3/2+Gamma} from \eqref{eq:3/2-}, \eqref{eq:3/2-Gam}
by replacing $\bar u$ by $-i \bar u \gamma_5$ for the spinor of the Y hyperon. The interesting aspect is that $-i \bar u \gamma_5$
satisfies the same equation of motion as $\bar u$ but with the mass replaced by its negative. As a consequence, most of the 
relations that we present now for the transition $3/2^{\pm} \to 1/2^{\pm}$ can be obtained from the corresponding relations 
for the transition $3/2^{\mp} \to 1/2^{\pm}$ from subsection \ref{sec:opp-par3/2} by just 
replacing $m_Y \to -m_Y$ (and changing $H \to \tilde H$). 

Again we introduce dimensionless helicity amplitudes:
\begin{eqnarray}
  \tilde H_-(q^2) &:=& - \left(m_{Y^*}+m_{Y}\right) m_{Y^*} \, \tilde H_1(q^2)  \nonumber \\
  && {}+\frac{1}{2} \left(m_{Y^*}^2-m_{Y}^2+q^2\right) \tilde H_2(q^2)+ q^2 \tilde H_3(q^2) \,, \nonumber \\
  \tilde H_0(q^2) &:=& -\left(m_{Y^*}+m_{Y}\right) m_{Y^*} \, \tilde H_1(q^2)   \nonumber \\
  && {} +\left(m_{Y^*}+m_{Y}\right) m_{Y^*} \, \tilde H_2(q^2)  
  \label{eq:defhelamp32+}  \\
  && {} + \frac{m_{Y^*}+m_{Y}}{2 m_{Y^*}} \left(m_{Y^*}^2-m_{Y}^2+q^2\right) \, \tilde H_3(q^2)  \,, 
    \nonumber \\
  \tilde H_+(q^2) &:=& -\left(q^2-m_{Y^*} m_{Y}-m_{Y}^2\right) \tilde H_1(q^2)     \nonumber \\
  && {} +\frac{1}{2} \left(m_{Y^*}^2-m_{Y}^2+q^2\right) \tilde H_2(q^2) + q^2 \tilde H_3(q^2)   \,.   \nonumber 
\end{eqnarray}
Note that the conventions for the helicity amplitudes are in line with \cite{Carlson:1985mm}, but 
opposite to \cite{Junker:2019vvy}. 

The kinematical constraints obtain the form
\begin{equation}
    \begin{split}
        \qquad\tilde H_+((m_{Y^*}+m_{Y})^2) &=  \tilde H_0((m_{Y^*}+m_{Y})^2)  \\ &=  \tilde H_-((m_{Y^*}+m_{Y})^2) 
    \end{split}
  \label{eq:helconstp3/2}
\end{equation}
and
\begin{equation}
    \begin{split}
        \qquad&\frac{2(m_{Y^*}-m_{Y})}{m_{Y^*}+m_{Y}} \tilde H_0((m_{Y^*}-m_{Y})^2)  \\ 
  &= \tilde H_+((m_{Y^*}-m_{Y})^2) + \tilde H_-((m_{Y^*}-m_{Y})^2)  \,.
    \end{split}
  \label{eq:helconstp3/2-low}
\end{equation}
The width for the two-body radiative decay $Y^* \to Y \gamma$ is given by
\begin{multline} 
    \Gamma_2 = e^2 \left[3| \tilde H_-(0)|^2 + | \tilde H_+(0)|^2 \right]  \\ 
    \times \frac{\left(m_{Y^*}-m_{Y}\right)^2 \left(m_{Y^*}^2-m_{Y}^2\right)}{96 \pi  m_{Y^*}^3}%
    \,.
    \label{eq:radiative32+}
\end{multline}

The differential decay width for the Dalitz decay $Y^* \to Y \gamma^* \to Y \, e^+ e^-$ can be expressed as
\begin{equation}
    \scalebox{0.96}{%
        $\begin{split} 
            &\frac{\text{d}\Gamma_3}{\text{d}q^2\text{d}(\cos{\theta})} = \frac{e^4 p_z \sqrt{q^2} \beta_e}{(2\pi)^3  \, 192 m_{Y^*}^3}
            \frac{(m_{Y^*}-m_{Y})^2-q^2}{q^2}  \\
            & \ \times\bigg\{\bigg(1+ \cos ^2\theta+ \frac{4 m_e^2}{q^2} \sin ^2\theta\bigg) \big[3 |\tilde H_-(q^2)|^2 + |\tilde H_+(q^2)|^2 \big] \\
            &\quad +\bigg(\sin ^2\theta+ \frac{4 m_e^2}{q^2}\cos ^2\theta\bigg)\frac{4q^2}{(m_{Y^*}+m_{Y})^2} |\tilde H_0(q^2)|^2\bigg\} \, .
        \end{split}$}
        \label{eq:dalitz32+}
\end{equation}
The four-body decay $Y^* \to Y \gamma^* \to \pi N \, e^+ e^- $ has the following differential decay width:
\begin{equation} 
    \scalebox{0.96}{%
        $\begin{split}
            &\frac{\text{d}\Gamma_4}{\text{d}q^2\text{d}(\cos{\theta})\text{d}\Omega_N}  \\
            &= \frac{e^4 p_z \sqrt{q^2} \beta_e }{(2\pi)^4 \, 384 m_{Y^*}^3} \, \text{Br}_{Y \to \pi N} 
            \, \frac{(m_{Y^*}-m_{Y})^2-q^2}{q^2} \\
            & \hspace{1mm} \times \bigg\{\bigg(1+ \cos ^2\theta+ \frac{4 m_e^2}{q^2} \sin ^2\theta\bigg) \big[|\tilde H_+(q^2)|^2 + 3 |\tilde H_-(q^2)|^2 \big] \\
            &\hspace{3mm}  +\bigg(\sin^2\theta+ \frac{4 m_e^2}{q^2}\cos^2\theta\bigg)
            \frac{4q^2}{\left(m_{Y^*}+m_{Y}\right)^2} |\tilde H_0(q^2)|^2    \\
            &\hspace{3mm}  +\frac{4\sqrt{q^2}\beta_e^2 }{m_{Y^*}+m_{Y}} \, \alpha _{Y} \, 
            \text{Im}[\tilde H_0(q^2) \tilde H_+^*(q^2)] \\
            &\qquad \times \sin \theta \cos \theta \sin \theta_N \sin \phi_N \bigg\}   \,.
        \end{split}$}
        \label{eq:fourbody-32+} 
\end{equation}

In the last equation we have a notable exception to the ``rule'' that the formulae of the present subsection are obtained 
from the corresponding ones in subsection \ref{sec:opp-par3/2} by the replacements $m_Y \to -m_Y$ and $H \to \tilde H$. 
The interference term has the opposite sign to the one in \eqref{eq:fourbody-32-}.

\subsection{Transition $\frac{1}{2}^{\pm} \to \frac{1}{2}^{\pm}$}
\label{sec:same-par1/2}

Finally we study the case of a transition between two hyperons with the same spin and parity assignments. One such case, the 
transition from $\Sigma^0$ to $\Lambda$ has been studied in detail in \cite{Granados:2017cib,Husek:2019wmt}. 

The decomposition is given by
\begin{equation}
    \resizebox{0.9\hsize}{!}{%
    $\langle p_Y,\lambda_Y \vert j^\mu(0) \vert p_{Y^*},\lambda_{Y^*} \rangle = 
    e \bar{u}(p_{Y},\lambda_{Y}) \, \Gamma^{\mu}_+ \, u(p_{Y^*},\lambda_{Y^*})
    \label{eq:1/2+}$%
    }
\end{equation}
with
\begin{equation}
     \Gamma^\mu_+ = i F_2(q^2) \, m_{Y^*} \, \sigma^{\mu\beta} q_\beta + F_3(q^2) \left(q^2\gamma^\mu - \slashed{q}q^\mu \right)  \,.
\end{equation}
We introduce the dimensionless helicity amplitudes by
\begin{equation}
    \begin{split}
        \qquad F_0(q^2) :=& (m_{Y^*}+m_{Y})^2 F_3(q^2)  \\ 
  &- (m_{Y^*}+m_{Y}) \, m_{Y^*} \, F_2(q^2)  \,, \\ 
  F_+(q^2) :=& q^2 F_3(q^2)- (m_{Y^*}+m_{Y})\, m_{Y^*} \, F_2(q^2)  \,.  
    \end{split}
  \label{eq:defhelp1/2}
\end{equation}
These helicity amplitudes satisfy the kinematical constraint 
\begin{equation}
  \qquad F_+((m_{Y^*}+m_{Y})^2) = F_0((m_{Y^*}+m_{Y})^2)  \,.
  \label{eq:helconstp1/2}
\end{equation}

The respective decay widths for the two-body radiative decay $Y^* \to Y \gamma$, 
the three-body Dalitz decay $Y^* \to Y \gamma^* \to Y \, e^+ e^-$,
and the four-body decay $Y^* \to Y \gamma^* \to \pi N \, e^+ e^-$ are given by
\begin{equation}
  \quad \Gamma_2 = \frac{e^2 |F_+(0)|^2 \left(m_{Y^*}-m_{Y}\right)^2 \left(m_{Y^*}^2-m_{Y}^2\right)}{8 \pi  m_{Y^*}^3}  \,,
  \label{eq:real1/2+}
\end{equation}
\begin{equation}
    \scalebox{0.98}{$\begin{split}
        &\frac{\text{d}\Gamma_3}{\text{d}q^2\text{d}(\cos{\theta})} = 
        \frac{e^4 p_z \sqrt{q^2}\beta_e}{(2\pi)^3 \, 16 m_{Y^*}^3}\frac{(m_{Y^*}-m_{Y})^2-q^2}{q^2} \\
        &\hspace{3mm} \times \bigg\{\bigg(1+\cos^2{\theta}+\frac{4m_e^2}{q^2}\sin^2{\theta} \bigg) \, |F_+(q^2)|^2 \\
        & \hspace{4mm} +\bigg(\sin^2{\theta}+\frac{4m_e^2}{q^2}\cos^2{\theta} \bigg)\frac{q^2}{(m_{Y^*}+m_{Y})^2}|F_0(q^2)|^2\bigg\}  \,,
    \end{split}$}
    \label{eq:12pddiff}
\end{equation}
and 
\begin{equation}
    \scalebox{0.96}{%
        $\begin{split}
            & \frac{\text{d}\Gamma_4}{\text{d}q^2\text{d}(\cos{\theta})\text{d}\Omega_N} \\
            & = \frac{e^4 p_z \sqrt{q^2} \beta _e}{(2\pi)^4 \, 32 m_{Y^*}^3} \text{Br}_{Y \to \pi N} 
            \, \frac{(m_{Y^*}-m_{Y})^2-q^2}{q^2} \\
            &\qquad \times \bigg\{\bigg(1+ \cos ^2\theta+ \frac{4 m_e^2}{q^2} \sin ^2\theta\bigg) |F_+(q^2)|^2  \\
            & \qquad \quad +\bigg(\sin ^2\theta+ \frac{4 m_e^2}{q^2}\cos ^2\theta\bigg) 
            \frac{q^2}{(m_{Y^*}+m_{Y})^2} |F_0(q^2)|^2 \\
            &\qquad \quad-\frac{2\sqrt{q^2} \beta_e^2}{m_{Y^*}+m_{Y}} \, \alpha _{Y} \, 
            \text{Im}[F_0(q^2) F_+^*(q^2)] \\
            &\qquad\qquad\times \sin \theta \cos \theta \sin \theta_N \sin \phi_N  \bigg\}   \,.
        \end{split}$}
        \label{eq:weak12p}
\end{equation}

\section{QED type case and form factor parametrization}
  \label{sec:FFparam}

In this section we will give a quantitative meaning to the phrases ``structureless'' and ``extended structure''. We will 
call the former ``QED type''. For the latter we introduce a radius. 

A real photon does not resolve the intrinsic structure of the composite hyperons that take part in the electromagnetic 
transitions. It is the photon virtuality that relates to the resolution; see e.g.\ the discussion 
in \cite{Alarcon:2017asr,Miller:2007uy}. In this spirit, we define a QED type case \cite{Junker:2019vvy}
by modifying the Dalitz decay formula \eqref{eq:dalitz32-} such that it fits to the radiative decay 
formula \eqref{eq:radiative32-}. To this end we replace e.g.\ in \eqref{eq:dalitz32-}
\begin{align}
    \big[3 |H_-(q^2)|^2 + |H_+(q^2)|^2 \big] &\to \big[3 |H_-(0)|^2 + |H_+(0)|^2 \big] \,, \nonumber \\
    \frac{4q^2}{(m_{Y^*}-m_{Y})^2} \, &|H_0(q^2)|^2  \to 0  \,,
    \label{eq:replQED}
\end{align}
i.e.\ we replace all virtualities $q^2$ by $0$ for these building blocks. We can do this for all spin-parity combinations. 
In this way we obtain a Dalitz decay formula for 
``structureless'' fermions. If we divide out the corresponding radiative decay width we get
\begin{eqnarray}
  && \frac{1}{\Gamma_2} \frac{\text{d}\Gamma_{\text{QED type}}}{\text{d}q^2\text{d}(\cos{\theta})} =    
  \frac{(m_{Y^*} \pm m_{Y})^2-q^2}{q^2 \left(m_{Y^*} \pm m_{Y}\right)^2 \left(m_{Y^*}^2-m_{Y}^2\right)}   \nonumber \\[1em]
  && \phantom{mmm} \times \frac{e^2 p_z \sqrt{q^2}\beta_e}{(4\pi)^2} 
  \bigg(1+\cos^2{\theta}+\frac{4m_e^2}{q^2}\sin^2{\theta} \bigg) \,.  \phantom{m} 
  \label{eq:QEDtypeDalitz}
\end{eqnarray}
Here the upper (lower) sign refers to the cases where the parities of initial and final fermion differ (are the same). 

Obviously, this formula \eqref{eq:QEDtypeDalitz} is independent of any transition form factors or helicity amplitudes. 
If an experiment cannot resolve the difference between nature and \eqref{eq:QEDtypeDalitz}, it cannot reveal if hyperons 
have an intrinsic structure or not. This does {\em not} mean that the determination of radiative decay widths does not 
contain any interesting information \cite{Kaxiras:1985zv}, but no visible deviation from \eqref{eq:QEDtypeDalitz} means 
that the such measured Dalitz decays do not contain more information than radiative decays, i.e.\ decays with real photons.

The QED type case \eqref{eq:QEDtypeDalitz} 
defines our baseline to which we want to compare the case where hyperons do have an intrinsic structure. 
From now on we focus on decays of the two negative-parity resonances $\Lambda(1520)$ and $\Lambda(1405)$. 
The mass difference between the considered resonances and the 
ground-state hyperons is not very large. Consequently the energy range $\sqrt{q^2}$ (dilepton invariant mass) 
that is explored by the Dalitz decays is rather limited, 
it ranges from two times the electron mass up to the mass difference $m_{Y^*}-m_Y$. For a rough estimate of the importance 
of the $q^2$ dependence, we approximate any 
helicity amplitude $G(q^2)$ in the following way:
\begin{equation}
  \label{eq:F0rad}
  \qquad \qquad G(q^2) \approx G(0) \left( 1 + \frac16 q^2 \langle r^2 \rangle \right) 
\end{equation}
where we have introduced the radius via 
\begin{equation}
  \label{eq:defrad}
  \qquad \qquad\langle r^2 \rangle := \frac{6}{G(0)} \left. \frac{\text{d}G(q^2)}{\text{d}q^2} \right\vert_{q^2 =0}  \,.
\end{equation}
In view of the typical size of hadrons of about $1\,$fm, we assume
\begin{equation}
 \qquad \qquad \quad 0 \le \langle r^2 \rangle \le 1 \,\text{fm}^2 \approx 25 \, \text{GeV}^{-2}  \,.
  \label{eq:upperlimit}  
\end{equation}
In practice, this will provide us with a rough upper limit of the deviation of a differential decay rate from the corresponding
QED type case. 

We would like to stress that \eqref{eq:upperlimit} is not entirely accurate. As already pointed out, the transition form factors 
and helicity amplitudes for resonances are not real-valued in the Dalitz decay region.\footnote{They are not even real-valued 
in the space-like region of electron-hadron scattering.} Strictly speaking, this carries over to the squared radii. The physical 
reason for being complex is the inelastic two-step process of strong decay, $Y^* \to \pi Y'$, and 
rescattering, $\pi Y' \to \gamma^* Y$. Here $Y'$ denotes another hyperon. However, the imaginary part of a transition form factor
or helicity amplitude will not be very large if the total decay width of the resonance is sufficiently small. This is the case 
for the resonances that we consider. For the electromagnetic transitions $\Sigma(1385) \to \Lambda$ the imaginary 
parts of squared radii have been 
explicitly calculated in \cite{Junker:2019vvy}. There, the real parts were in the ballpark of \eqref{eq:upperlimit} and 
the imaginary parts were much smaller. Therefore we assume that the real part of any $\langle r^2 \rangle$ is dominant over the 
imaginary 
part. 

What remains to be shown to justify \eqref{eq:upperlimit} as a reasonable approximation? We still have to argue why 
the real part should have positive sign. 
Indeed, there is a well-known case where the squared radius of a hadron seems to be negative: 
the electric charge radius of the neutron \cite{pdg}. Yet, this is a somewhat misleading case. For the electric form factor of 
the neutron, the radius cannot be defined via \eqref{eq:defrad}, since the charge of the neutron, $G(0)$, vanishes. As a remedy
one drops $G(0)$ in the definition of the charge radius of the neutron. However, one can take a somewhat different route 
to the same physical information. Instead of electric charge, one can look at the isospin. The isoscalar and isovector form 
factors of the nucleon have all non-vanishing ``charges''. One can define an isoscalar and an isovector radius based on 
\eqref{eq:defrad}, see, e.g., \cite{Leupold:2017ngs,Kubis:2000aa}. 
Those two radii are positive and in the order of 1 fm, in full agreement with \eqref{eq:upperlimit}. 
Also the results of \cite{Junker:2019vvy} support this estimate. Finally we note that even the radius of the pion transition 
form factor \cite{Hoferichter:2014vra} fits into this picture. For all these cases, the radii are less than 1 fm. Thus we 
believe that \eqref{eq:upperlimit} defines a reasonable conservative range and we expect that for all hadrons, reality is 
closer to 1 fm than to 0.

\section{Concrete results for negative-parity resonances}
\label{sec:concrete}

The parametrization \eqref{eq:F0rad} seems to indicate that we have two free parameters per helicity amplitude. However, this 
is not the case because we have to obey the kinematical constraints \eqref{eq:helconstm3/2} or \eqref{eq:helconstm1/2}, 
respectively. This will help us to express all relevant quantities solely in terms of radii. 

Some clarification is in order here. Of course, the ansatz \eqref{eq:F0rad} is an approximation that holds only for 
sufficiently low values of $q^2$. We want to use this approximation for the whole Dalitz decay region for cases where the mass 
difference between the decaying and the final hyperon is sufficiently small. The kinematical constraints 
\eqref{eq:helconstm3/2} or \eqref{eq:helconstm1/2} lie in this low-energy region because they lie at the end of the 
kinematically allowed Dalitz decay region. Therefore we can use the kinematical constraints to reduce the number of free 
parameters and focus on the impact of the radii on the results. Note that this line of reasoning would {\em not} work for 
positive-parity resonances. In particular, the constraint \eqref{eq:helconstp1/2} does {\em not} lie in the low-energy 
region where the ansatz \eqref{eq:F0rad} would make sense. Albeit not of completely general use, the kinematical constraints are 
absolutely suited for the cases that we consider further, namely for the hyperon resonances $\Lambda(1520)$ and $\Lambda(1405)$
with spin-parity assignment $3/2^-$ and $1/2^-$, respectively \cite{pdg}.

\subsection{Transition radii and the decay $\Lambda(1405) \to \Lambda \, e^+ e^-$}
\label{sec:1405}

Using \eqref{eq:F0rad} for the helicity amplitudes $\tilde F_+$ and $\tilde F_0$ allows us to rewrite the kinematical constraint \eqref{eq:helconstm1/2} into 
\begin{equation}
  \qquad \tilde F_0(0) \approx \tilde F_+(0) \, 
   \frac{1+ (m_{Y^*}-m_Y)^2 \langle r^2 \rangle_+/6}{1+ (m_{Y^*}-m_Y)^2 \langle r^2 \rangle_0/6}  \,.
   \label{eq:const12}
\end{equation}
The remaining dependence on $\tilde F_+(0)$ cancels out in the ratio of the Dalitz decay width \eqref{eq:12mddiff} and 
the radiative decay width \eqref{eq:real1/2-}. Thus the such normalized decay width 
$ \text{d}\Gamma_3/(\text{d}q^2 \text{d}(\cos\theta))/\Gamma_2$ depends only on the (squared) transition radii 
$\langle r^2 \rangle_+$ and $\langle r^2 \rangle_0$. 

\begin{figure}[!ht]
  \centering
  \includegraphics[width=0.4\textwidth]{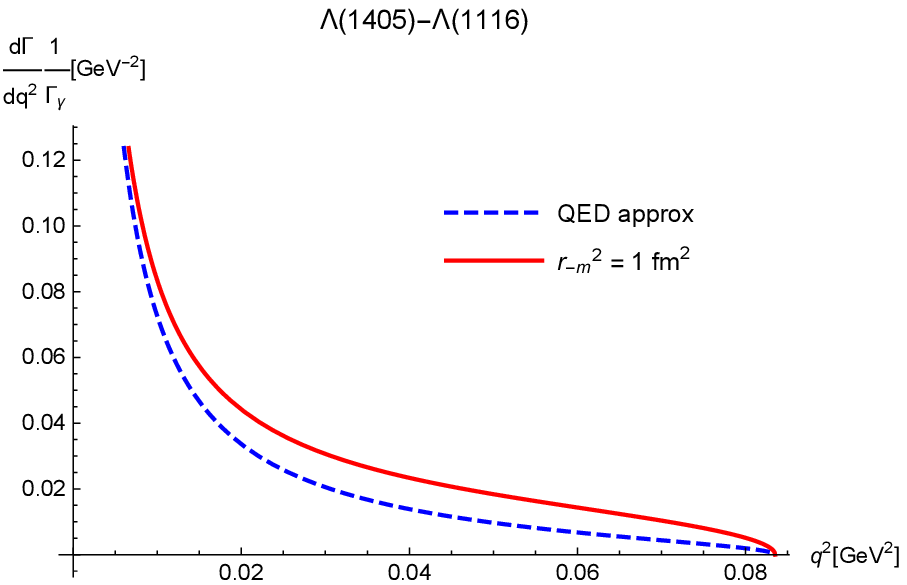}
  \vspace{1mm}
  \includegraphics[width=0.4\textwidth]{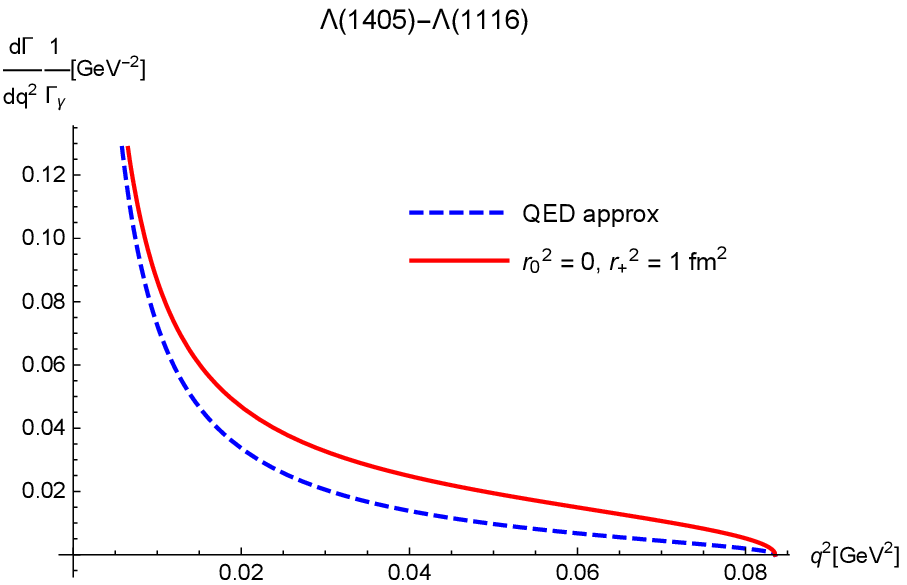}
  \vspace{1mm}
  \includegraphics[width=0.4\textwidth]{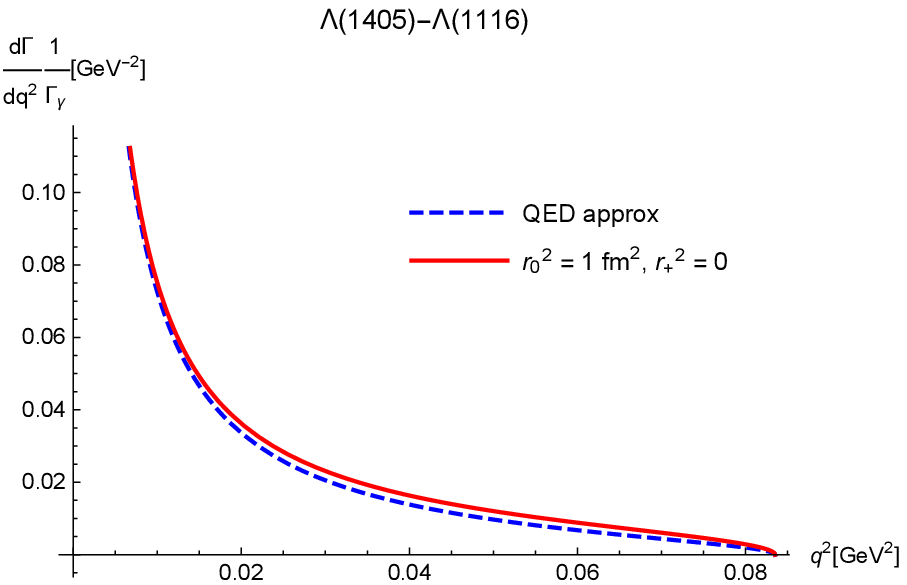}
  \vspace{1mm}  
  \includegraphics[width=0.4\textwidth]{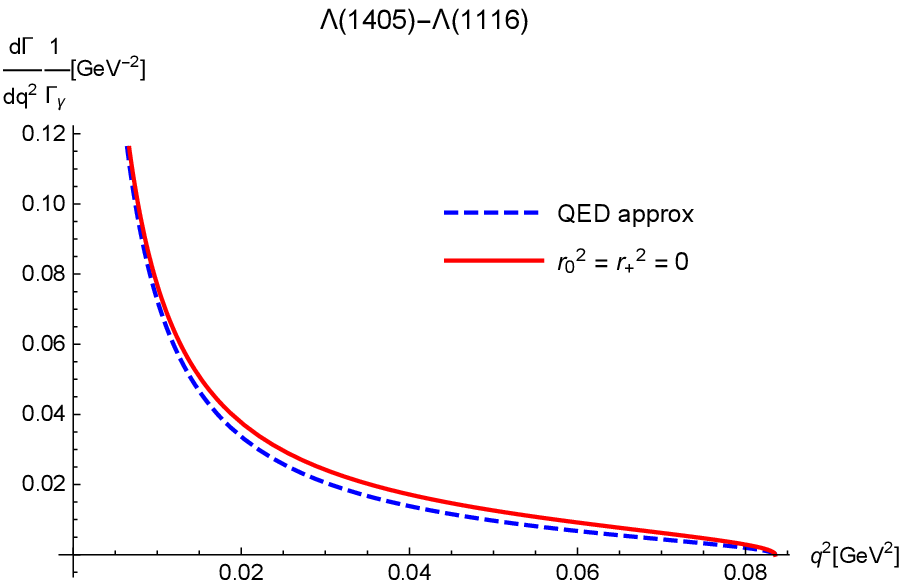}
  \caption{Comparison between radius structure and QED-type approximation for the $\frac{1}{2}^- \to \frac{1}{2}^+$ 
    transition.
    {\it Top panel:} both radii at maximal value \eqref{eq:upperlimit}; {\it middle panels:} one radius at max, one zero; 
    {\it bottom panel:} both radii put to zero.}
  \label{fig:radius12}
\end{figure}
In the following, we will focus on the once-integrated normalized decay widths  
\begin{equation}
  \qquad \qquad \qquad \label{eq:singly-decayw-12m}
  \frac{1}{\Gamma_2} \frac{\text{d}\Gamma_3}{\text{d}q^2} \quad \mbox{and} \quad
  \frac{1}{\Gamma_2} \frac{\text{d}\Gamma_3}{\text{d}(\cos\theta)} \,,
\end{equation}
to study the dependence on the transition radii and compare to the QED type case.

The $q^2$ dependence is depicted in fig.\ \ref{fig:radius12}.
We observe that the radius related to the helicity flip amplitude $\tilde F_+$ causes a significant deviation from the 
QED type case. The effect from the non-flip amplitude is minor. This is related to the additional $q^2$ factor that 
multiplies $\tilde F_0$ in \eqref{eq:12mddiff} and to the larger weight of $1+\cos^2\theta$ relative to 
$\sin^2\theta = 1 - \cos^2\theta$ when integrating over $\cos\theta$. 
Note that the QED type case is {\em not} defined by putting all radii to zero.
Therefore, there is a small difference between the two curves in the bottom panel of fig.\ \ref{fig:radius12}. 
We could have defined the QED type case in a different way. But instead we use the occasion to point out that there is in 
principle an ambiguity in defining the structureless QED case. In practice, however, this ambiguity is small. 

Depending on the composition of the $\Lambda(1405)$ as a dominantly three-quark or dominantly hadron molecule state, 
its helicity flip transition radius can be expected to be somewhat different. Still we think that 1 fm is a reasonable 
estimate in any case. We regard the plots of fig.\ \ref{fig:radius12} as interesting for experimentalists who aim to reveal the 
intrinsic structure of the $\Lambda(1405)$ using Dalitz decays. For this endeavor one needs to achieve an experimental 
accuracy that can discriminate at least between the two lines in the top panel of fig.\ \ref{fig:radius12}. To study 
differences between different scenarios for the structure of the $\Lambda(1405)$ requires an even better accuracy.

Next we turn to the angular distribution. If one integrates \eqref{eq:12mddiff} or \eqref{eq:QEDtypeDalitz} over $q^2$ 
(in the range $(2m_e)^2 \le q^2 \le (m_{Y^*}-m_Y)^2$), one obtains a constant term and one linear in $\cos^2\theta$. Thus 
the general structure is 
\begin{equation}
  \label{eq:singly-decayw-12m-angles}
  \qquad \qquad\qquad\frac{1}{\Gamma_2} \frac{\text{d}\Gamma_3}{\text{d}(\cos\theta)} = A \, (\cos^2\theta + C)  \,.
\end{equation}
If one integrates finally over the angle, one obtains the total decay width for the process $Y^* \to Y \, e^+ e^-$, normalized 
to the width for the process $Y^* \to Y \gamma$, i.e.
\begin{equation}
  \label{eq:total-decayw-12m}
  \qquad \qquad\qquad \qquad\frac{\Gamma_3}{\Gamma_2} = \frac23 \, A + 2 A C  \,. 
\end{equation}
In table \ref{table:ABCvalues12} we provide $A$, $C$ and the width ratio of \eqref{eq:total-decayw-12m} 
for the QED type case and different radius combinations. 
\begin{table}[!ht]
  \centering
  \begin{tabular}[t]{cccc}
    \hline
    $\big(\langle r^2 \rangle_0; \langle r^2 \rangle_+\big)$ GeV$^{-2}$ &  $A$ & $C$ & $\frac{2}{3}A + 2AC$  \\
    \hline
    \big(0 ; 25\big) & $0.00252$ & 1.43 & 0.00889 \\[0.3em]
    \big(25 ; 25\big) & $0.00262$ & 1.34 & 0.00876\\[0.3em]
    \big(0 ; 0\big) & $0.00253$ & 1.30 & 0.00830 \\[0.3em]
    \big(25 ; 0\big) & $0.00259$ & 1.26 & 0.00823 \\
    \hline
    \hline
    QED type & $0.00273$ & 1.14 & 0.00804 \\
    \hline
 \end{tabular}
  \caption{Parameters $A$ and $C$, and decay width ratio for the $J^P=\frac{1}{2}^-$ initial state in QED approximation 
    and with a radius structure.}
  \label{table:ABCvalues12}
\end{table}
Concerning the overall scaling factor $A$, we observe that different radii lead to modifications on a 10\% level. However, 
for this parameter the ambiguity how to define a structureless case is in the same order.

For the parameter $C$, larger 
helicity flip transition radii lead to larger values. In total, we observe variations up to about 20\%. Interestingly, a radius $\langle r^2 \rangle_0$ in the non-flip amplitude counteracts the effect of a radius $\langle r^2 \rangle_+$ in the helicity flip amplitude.

We predict that the Dalitz decay width, normalized to the radiative decay width, is about 0.8 to 0.9\%. This fits to the rule of thumb that an additional QED vertex provides a suppression factor of about $\alpha = e^2/(4\pi) \approx 10^{-2}$.

\subsection{Transition radii and the decay $\Lambda(1520) \to \Lambda \, e^+ e^-$}
\label{sec:1520}

\begin{figure}[!ht]
    \centering
    \includegraphics[width=0.4\textwidth]{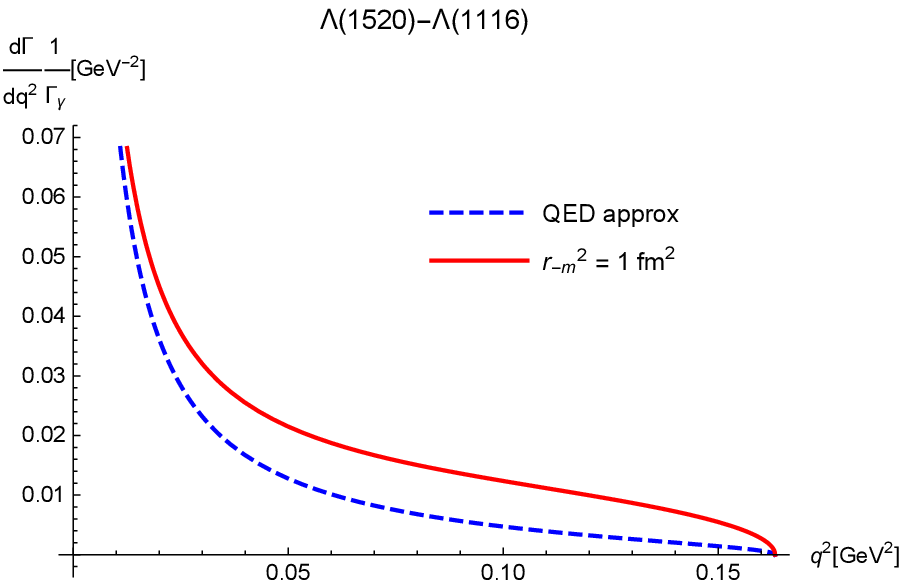}
    \vspace{1mm}
    \includegraphics[width=0.4\textwidth]{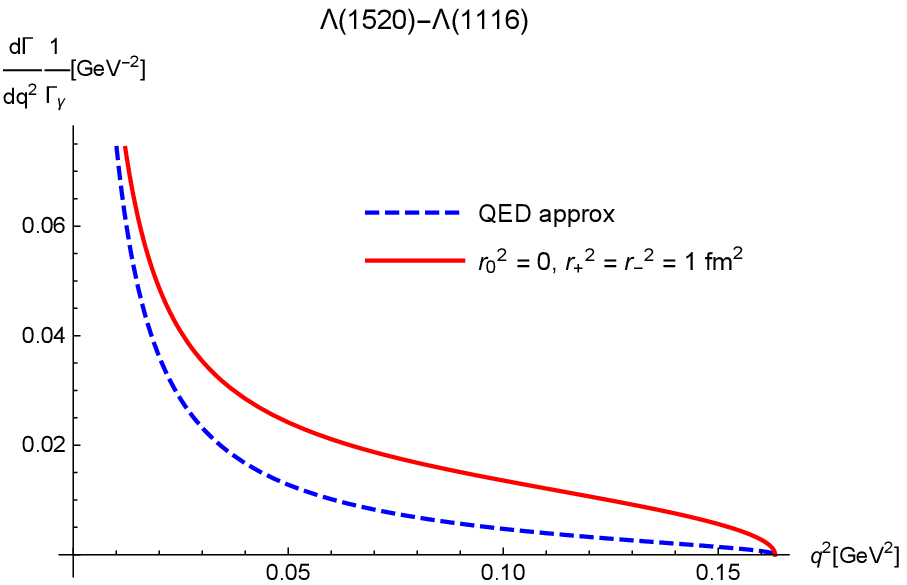}
    \vspace{1mm}
    \includegraphics[width=0.4\textwidth]{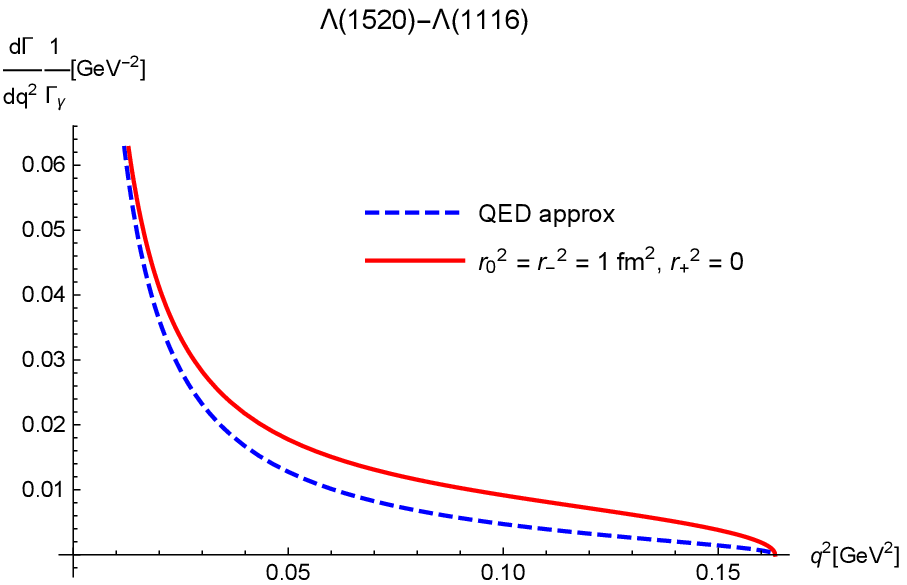}
    \vspace{1mm}
    \includegraphics[width=0.4\textwidth]{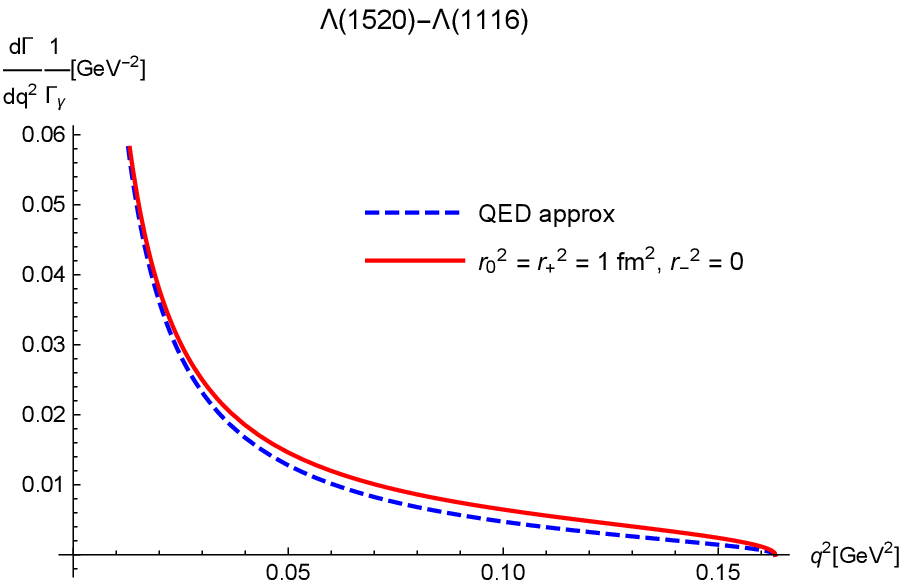}
    \caption{Same as fig.\ \ref{fig:radius12} but for the $\frac{3}{2}^- \to \frac{1}{2}^+$ 
      transition.
      {\it Top panel:} all three radii at maximal value \eqref{eq:upperlimit}; {\it other panels:} two radii at max, third at zero.}
    \label{fig:radius32-twonon0}
\end{figure}

\begin{figure}[!ht]
    \centering
    \includegraphics[width=0.4\textwidth]{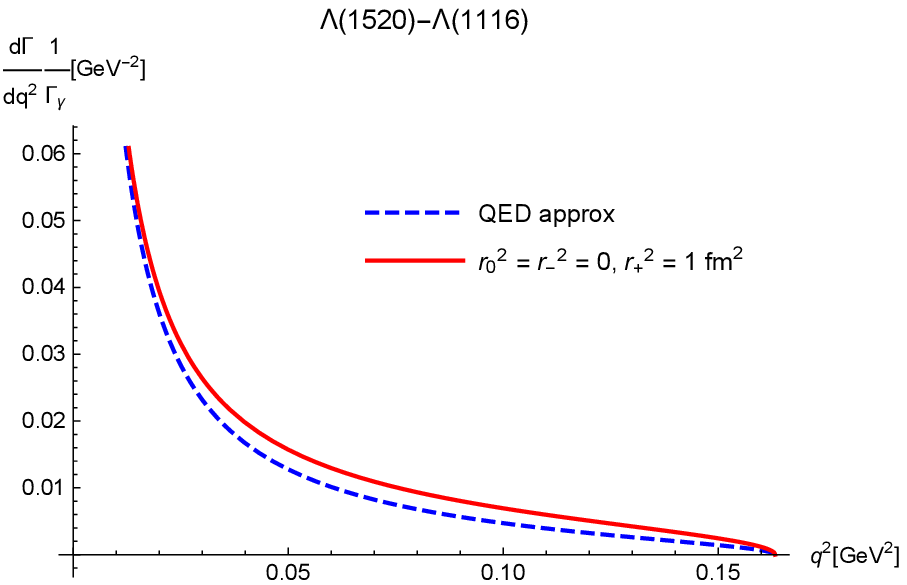}
    \vspace{1mm}
    \includegraphics[width=0.4\textwidth]{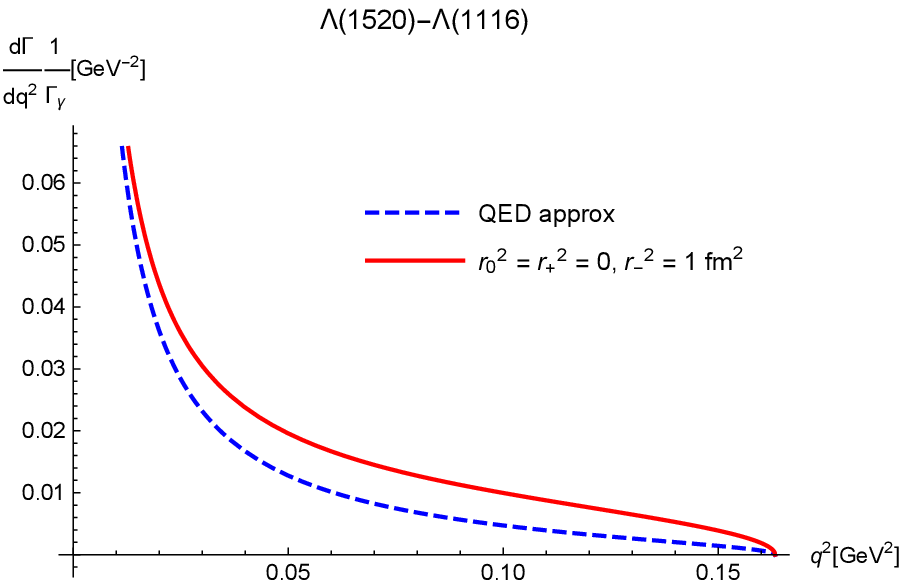}
    \vspace{1mm}
    \includegraphics[width=0.4\textwidth]{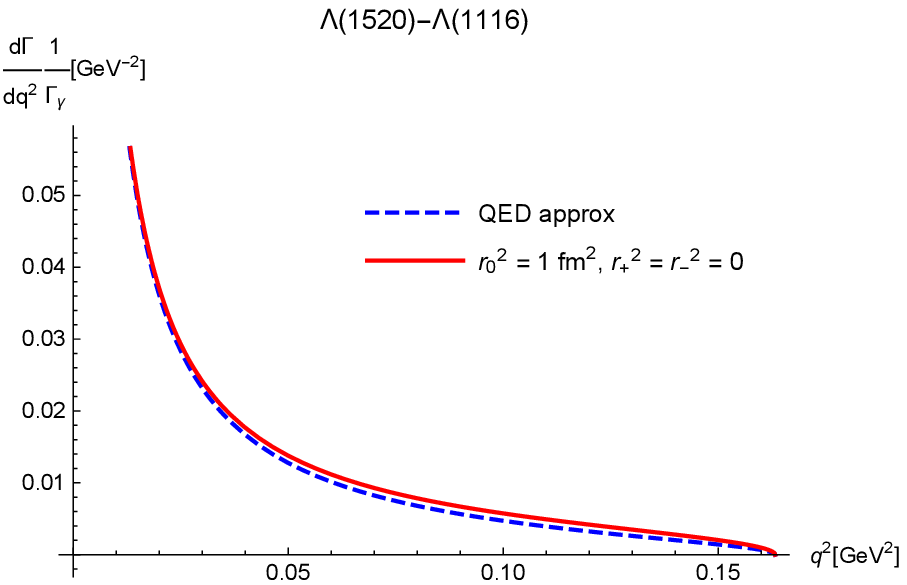}
    \vspace{1mm}
    \includegraphics[width=0.4\textwidth]{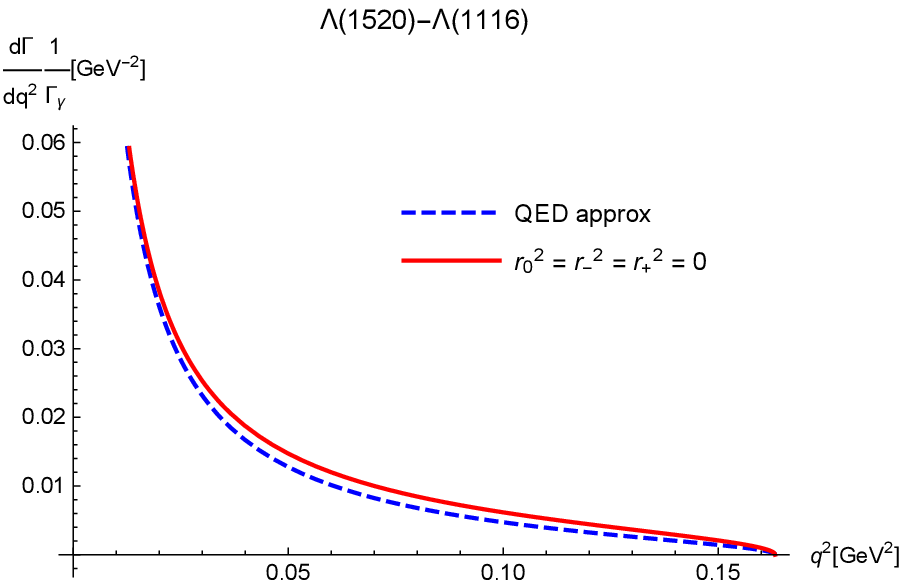}
    \caption{Same as fig.\ \ref{fig:radius32-twonon0} but with less many radii put at maximum value \eqref{eq:upperlimit}. 
      {\it Bottom panel:} all radii put to zero; {\it other panels:} one radius at max, rest at zero.}
    \label{fig:radius32-onenon0}
\end{figure}

We perform the same procedure for the decays of the spin-$3/2^-$ hyperon resonance $Y^* = \Lambda(1520)$. \\For the numerical results we focus on $Y = \Lambda$, the lightest strange baryon. The approximation 
\eqref{eq:F0rad} is utilized for the helicity amplitudes $H_+$, $H_0$, $H_-$. It is supposed to be valid in the whole Dalitz decay region, $(2m_e)^2 \le q^2 \le (m_{Y^*}-m_Y)^2$. The kinematical constraints \eqref{eq:helconstm3/2} are used to eliminate the dependence on $H_{...}(0)$ for the ratio of Dalitz decay width and radiative width. What remains is the dependence on the three squared radii, $\langle r^2 \rangle_+$, $\langle r^2 \rangle_0$, $\langle r^2 \rangle_-$. 
Figs.\ \ref{fig:radius32-twonon0} and \ref{fig:radius32-onenon0} illustrate this dependence as a function of $q^2$ for the normalized singly-differential Dalitz decay width.
What the plots show, first of all, is the fact that $\langle r^2 \rangle_-$ matters most. Whenever it is large, there is a significant deviation from the QED type case. Whenever it is small, the results with radii are close to the structureless QED case. Essentially this can be traced back to the explicit factor of 3 in \eqref{eq:dalitz32-} that boosts the importance of $H_-$ relative to the other two helicity amplitudes. In addition, the importance of $\vert H_0 \vert^2$ is demoted by the explicit $q^2$ factor in front of it and the different weight in the angular average, an effect already observed for the spin-1/2 case of the previous subsection.  
Again, we believe that the plots of figs.\ \ref{fig:radius32-twonon0} and \ref{fig:radius32-onenon0} are important for experimentalists to judge how accurate their results must be to disentangle extended from structureless objects. 

We turn to the angular dependence. Formulae \eqref{eq:singly-decayw-12m-angles} and \eqref{eq:total-decayw-12m} can again be used and the dependence of the parameters $A$ and $C$ on the radii is collected in table \ref{table:ABCvalues32}.
\begin{table}[!ht]
  \centering
  \begin{tabular}[t]{cccc}
    \hline
    $\big(\langle r^2 \rangle_0; \langle r^2 \rangle_+; \langle r^2 \rangle_-\big)$ & \multirow[c]{2}{*}{$A$} & \multirow[c]{2}{*}{$C$} & \multirow[c]{2}{*}{$\frac{2}{3}A + 2AC$}  \\ GeV$^{-2}$ & & & \\ 
    \hline
    \big(0 ; 25 ; 25\big) & $0.00267$ & 1.55 & 0.0101  \\[0.3em]
    \big(0 ; 0 ; 25\big) & $0.00270$ & 1.42 & 0.00945  \\[0.3em]
    \big(25 ; 25 ; 25\big) & $0.00290$ & 1.35 & 0.00976  \\[0.3em]
    \big(0 ; 25 ; 0\big) & $0.00272$ & 1.31 & 0.00894  \\[0.3em]
    \big(25 ; 0 ; 25\big) & $0.00285$ & 1.29 & 0.00924  \\[0.3em]
    \big(0 ; 0 ; 0\big) & $0.00273$ & 1.28 & 0.00881  \\[0.3em]
    \big(25 ; 25 ; 0\big) & $0.00282$ & 1.23 & 0.00881  \\[0.3em]
    \big(25 ; 0 ; 0\big) & $0.00281$ & 1.22 & 0.00870   \\
    \hline
    \hline
    QED type & $0.00292$ & 1.13 & 0.00855 \\
    \hline
  \end{tabular}
  \caption{Same as table \ref{table:ABCvalues12} but for the $J^P=\frac{3}{2}^-$ initial state.}
  \label{table:ABCvalues32}
\end{table}
Similar to the spin-1/2 case of the previous subsection we observe a variation in $A$ of about 10\%, but no clear tendency of 
the impact of finite radii.

Concerning the parameter $C$ we observe variations of up to 30\%. 
The value of $C$ is increased for larger 
values of $\langle r^2 \rangle_-$ or $\langle r^2 \rangle_+$ and decreased for a larger value of $\langle r^2 \rangle_0$. 
The impact of $\langle r^2 \rangle_+$ is less pronounced than the impact of $\langle r^2 \rangle_-$, which we explain again 
by the explicit factor of 3 in \eqref{eq:dalitz32-}. 

For the integrated Dalitz decay width, normalized to the photon decay width, we predict 0.9 to 1\%, slightly larger 
than our prediction for the $\Lambda(1405)$ of the previous subsection. We attribute this to the larger phase space available 
for the heavier $\Lambda(1520)$.

\section{Further discussion, summary and outlook}
\label{sec:summary}

We have provided a comprehensive framework that first considers the most general electromagnetic transition form factors, free of 
kinematical constraints. Those form factors are perfectly suited for a dispersive representation \cite{Junker:2019vvy}, 
a perspective that we might explore in the future. We have related these form factors to helicity amplitudes and shown how they make 
their appearance in the decay rates of the processes $Y^* \to Y \gamma$ and $Y^* \to Y \, e^- e^+$. We covered the cases where 
the initial hyperon $Y^*$ can have spin 1/2 or 3/2. The final hyperon $Y$ was assumed to have spin 1/2. All parity combinations 
have been covered. 

Of course, these relations 
are in principle not new and easy to deduce, e.g., from \cite{Korner:1976hv}. Yet, we have decided to spell out all 
definitions and conventions in detail to keep the work self-contained and to facilitate comparisons between different works. 
We have further extended the framework to include also 
a final weak decay of the ground-state hyperon $Y$, very much in spirit of \cite{Perotti:2018wxm}, though with different 
kinematics. We have stressed that the 
emerging interference term between a helicity flip and the non-flip amplitude exists also in the Dalitz decay region. 

In a second step, we have focused on the specific transitions of $\Lambda(1405)$ and $\Lambda(1520)$ to the ground-state 
$\Lambda$. In view of the not too large kinematical range that is covered by the Dalitz decays we have parametrized the 
helicity amplitudes in terms of transition radii. Such radii are commonly used for the study of form factors at low energies, 
e.g.\ for the electric or magnetic form factors of proton and neutron, i.e.\ for their helicity non-flip or flip amplitudes. 
As a function of such transition radii, we have investigated how much accuracy is needed for an experiment to discriminate 
between hypothetical structureless hyperons and a more realistic case. Indeed, we regard the scenario where all 
radii are non-vanishing as the most realistic one. 

Concerning the dependence on the invariant mass of the electron-positron 
pair, our setup shows a significant deviation from the structureless QED type case, 
in particular in the $q^2$ range above $0.01\,$GeV$^2$ for the transition from $\Lambda(1405)$ to $\Lambda$ and above 
$0.02\,$GeV$^2$ for the transition from $\Lambda(1520)$ to $\Lambda$. We see also significant differences for the parameter $C$
that parametrizes the deviation of the angular dependence from a pure $\cos^2$ distribution. 

Of course, it is completely straightforward to extend this radius analysis also to the electromagnetic 
transitions $\Lambda(1405) \to \Sigma^0$, $\Lambda(1405) \to \Sigma^0(1385)$, $\Lambda(1520) \to \Sigma^0$, 
and with a little generalization also to $\Lambda(1520) \to \Sigma^0(1385)$. 
But the smaller the mass difference between initial and final hyperon, the less additional information is contained in the 
Dalitz decays relative to the real-photon case. Therefore we have focused on the cases where the final hyperon is as light 
as possible. 

More generally, the framework provided here can also be applied to double- or triple-strange hyperon 
resonances and to baryon resonances 
where strange quarks are replaced by heavier quarks.

\begin{acknowledgements}
SL thanks T.\ Galatyuk, A.\ Kup\'s\'c, B.\ Ramstein, P.\ Salabura and K.\ Sch\"onning for many valuable discussions that 
inspired this investigation. 
This work has been supported by the Swedish Research Council (Vetenskapsr\aa det) (grant number 2019-04303). 
Partial support has been provided by the Polish National Science Centre through the grant 2019/35/O/ST2/02907. 
\end{acknowledgements}

\appendix

\section{Toy-model Lagrangians}
\label{sec:effL}

The following Lagrangians are hermitian and invariant with respect to charge conjugation symmetry. They are also 
parity symmetric, but it is the first two properties that make the coupling constants real-valued. Note that we do not make use 
of these Lagrangians in our actual calculations. In this sense, these are toy-model Lagrangians. But we use them to motivate 
the appearance or absence of $i$'s for our definitions of the transition form factors $H_j$, $\tilde H_j$, $F_j$, 
and $\tilde F_j$. 

A real-valued contribution to $H_1$ as introduced in \eqref{eq:3/2-Gam} comes from a tree-level calculation based on
\begin{eqnarray}
  \label{eq:L4H1}
  \qquad\qquad{\cal L}_{H1} = a_1 \left(\bar\Psi \gamma^\mu F_{\mu\nu} \Psi^\nu + \bar\Psi^\nu \gamma^\mu F_{\mu\nu} \Psi \right) \,,
\end{eqnarray}
where $a_1 \in \mathds{R}$. 
In the same way one finds a contribution from
\begin{eqnarray}
  \label{eq:L4H2}
  \qquad\qquad{\cal L}_{H2} = i a_2 \left( \bar\Psi F_{\mu\nu} \partial^\mu \Psi^\nu - \partial^\mu \bar\Psi^\nu F_{\mu\nu} \Psi \right)
\end{eqnarray}
for $H_2$ and from
\begin{eqnarray}
  \label{eq:L4H3}
  \qquad\qquad {\cal L}_{H3} = i a_3 \left( \bar\Psi \partial^\mu F_{\mu\nu} \Psi^\nu - \bar\Psi^\nu \partial^\mu F_{\mu\nu} \Psi \right)
\end{eqnarray}
for $H_3$. Also here the coupling constants must be real, $a_2, a_3 \in \mathds{R}$.

Similarly, the Lagrangians
\begin{equation}
  \label{eq:L4tildeH1}
  \qquad\tilde {\cal L}_{H1} = i \tilde a_1 \left(\bar\Psi \gamma^\mu \gamma_5 F_{\mu\nu} \Psi^\nu 
    - \bar\Psi^\nu \gamma^\mu \gamma_5 F_{\mu\nu} \Psi \right) \,,
\end{equation}
\begin{equation}
  \label{eq:L4tildeH2}
  \qquad\tilde {\cal L}_{H2} = \tilde a_2 \left(\bar\Psi \gamma_5 F_{\mu\nu} \partial^\mu \Psi^\nu 
    - \partial^\mu \bar\Psi^\nu \gamma_5 F_{\mu\nu} \Psi \right)  \,,
\end{equation}
\begin{equation}
  \label{eq:L4tildeH3}
  \qquad\tilde {\cal L}_{H3} = \tilde a_3 \left(\bar\Psi \gamma_5 \partial^\mu F_{\mu\nu} \Psi^\nu 
    - \bar\Psi^\nu \gamma_5 \partial^\mu F_{\mu\nu} \Psi \right)
\end{equation}
contribute to $\tilde H_1$, $\tilde H_2$, $\tilde H_3$, respectively. We recall that the latter transition form factors have been 
introduced in \eqref{eq:3/2+Gamma}. The Lagrangians are only hermitian and invariant with respect to charge conjugation, if 
the coupling constants are real, i.e.\ $\tilde a_1 , \tilde a_2 , \tilde a_3 \in \mathds{R}$. 

Turning to spin-1/2, we have the Lagrangians
\begin{equation}
  \label{eq:L4F2}
  \qquad{\cal L}_{F2} = b_2 \left(\bar\Psi_{Y} \sigma_{\mu\nu} F^{\mu\nu} \Psi_{Y^*} 
  + \bar\Psi_{Y^*} \sigma_{\mu\nu} F^{\mu\nu} \Psi_{Y} \right)
\end{equation}
and
\begin{equation}
  \quad\label{eq:L4F3}
  {\cal L}_{F3} = b_3 \left(\bar\Psi_{Y} \gamma_\mu \partial_\nu F^{\mu\nu} \Psi_{Y^*} 
  + \bar\Psi_{Y^*} \gamma_\mu \partial_\nu F^{\mu\nu} \Psi_{Y} \right)
\end{equation}
contributing with real-valued results to $F_2$ and $F_3$, respectively. Finally
\begin{equation}
  \label{eq:L4tildeF2}
  \scalebox{0.99}{$\tilde {\cal L}_{F2} = i \tilde b_2 \left(\bar\Psi_{Y} \sigma_{\mu\nu} \gamma_5 F^{\mu\nu} \Psi_{Y^*} 
  + \bar\Psi_{Y^*} \sigma_{\mu\nu} \gamma_5 F^{\mu\nu} \Psi_{Y} \right)$}
\end{equation}
and
\begin{equation}
\resizebox{0.86\hsize}{!}{%
  $\label{eq:L4tildeF3}
  \tilde {\cal L}_{F3} = i \tilde b_3 \left(\bar\Psi_{Y} \gamma_\mu \gamma_5 \partial_\nu F^{\mu\nu} \Psi_{Y^*} 
  - \bar\Psi_{Y^*} \gamma_\mu \gamma_5 \partial_\nu F^{\mu\nu} \Psi_{Y} \right)$%
  }
\end{equation}
contribute to $\tilde F_2$ and $\tilde F_3$, respectively. Hermiticity and charge conjugation symmetry demand 
$b_2, b_3, \tilde b_2, \tilde b_3 \in \mathds{R}$. 

An explanation is in order concerning the labels 2 and 3 for the transition form factors $F_i$ and $\tilde F_i$. 
In a (here formal) low-energy counting scheme where the hyperons' three-momenta and 
the electromagnetic potentials and their four-momenta are counted as soft and of same size, the Lagrangians \eqref{eq:L4F2} 
and \eqref{eq:L4tildeF2} are of second order, while \eqref{eq:L4F3} and \eqref{eq:L4tildeF3} are of third order. 

In addition, the construction in \eqref{eq:L4F2} resembles the Pauli term that describes the anomalous magnetic moment of a 
spin-1/2 state, here generalized to transitions. For the nucleon, the Pauli form factor is often called $F_2$; 
see e.g.\ \cite{Kubis:2000aa} and references therein. 

For the transitions involving spin-3/2 initial states this formal power counting leads to a mismatch between the 
two parity sectors, cf.\ e.g.\ \cite{Holmberg:2019ltw,Junker:2019vvy}.
Therefore we have just enumerated the transition form factors $H_i$ and $\tilde H_i$ from 1 to 3.

\section{High-energy behavior}
\label{sec:high-qcr}

In this appendix we use the methods of \cite{Carlson:1985mm} to present the high-energy behavior of all transition 
form factors. This procedure uses the implicit assumption that the considered hyperons have a minimal quark content of three 
quarks. For the validity of the framework, it does not matter how large the overlap of the physical state with a 
three-quark configuration is \cite{Aznauryan:2011qj}, as long as it is non-zero. 

The quark counting rules are derived for large space-like photon virtuality. Therefore it is convenient to introduce the large
positive quantity $Q^2 := -q^2$ and $Q:=\sqrt{Q^2}$. One obtains 
\begin{equation}
    \begin{split}
        H_+(-Q^2) \sim \frac{1}{Q^4} \,, &\;   H_0(-Q^2) \sim \frac{1}{Q^6} \,, \;   H_-(-Q^2) \sim \frac{1}{Q^6} \,;  \\[1em]
        \tilde H_+(-Q^2) \sim \frac{1}{Q^4} \,, &\;   \tilde H_0(-Q^2) \sim \frac{1}{Q^6} \,, \;   
        \tilde H_-(-Q^2) \sim \frac{1}{Q^6} \,; \\[1em]
        F_+(-Q^2) \sim \frac{1}{Q^4} \,, &\;   F_0(-Q^2) \sim \frac{1}{Q^6} \,; \\[1em]
        \tilde F_+(-Q^2) \sim \frac{1}{Q^4} \,, &\;   \tilde F_0(-Q^2) \sim \frac{1}{Q^6} \,,
    \end{split}
    \label{eq:high-en-scaling}  
\end{equation}
which leads to 
\begin{equation}
    \begin{split}
        H_1(-Q^2) \sim \frac{1}{Q^6} \,, &\;   H_2(-Q^2) \sim \frac{1}{Q^8} \,, \;   H_3(-Q^2) \sim \frac{1}{Q^8} \,;  \\[1em]
        \tilde H_1(-Q^2) \sim \frac{1}{Q^6} \,, &\;  \tilde H_2(-Q^2) \sim \frac{1}{Q^8} \,, \;   
        \tilde H_3(-Q^2) \sim \frac{1}{Q^8} \,;  \\[1em]
        F_2(-Q^2) \sim \frac{1}{Q^6} \,, &\;   F_3(-Q^2) \sim \frac{1}{Q^6} \,; \\[1em]
        \tilde F_2(-Q^2) \sim \frac{1}{Q^6} \,, &\;  \tilde F_3(-Q^2) \sim \frac{1}{Q^6} \,.
    \end{split}
    \label{eq:high-en-scaling2}  
\end{equation}

\section{Kinematical constraints and partial waves}
\label{sec:multi-p}

In the main part of this paper we use helicity amplitudes, i.e.\ amplitudes characterized by the helicities of the initial
and final states and by total angular momentum \cite{Jacob:1959at}. Instead of helicities, one could also use
spin and orbital angular
momentum. This characterization scheme is commonly used in non-relativistic physics (where the spin-orbit coupling
is suppressed). While we find the helicity amplitudes in general more practical for our purpose, the classification
using orbital angular momentum $l$ is actually helpful where the considered system becomes non-relativistic.
For the Dalitz decays $Y^* \to Y \, e^+ e^-$, this happens at the end of the phase space,
i.e.\ for $q^2 \approx (m_{Y^*}-m_Y)^2$. Here initial and final hyperon are at rest and their mass difference goes entirely into
the back-to-back motion of the electron-positron pair. Note that we consider the frame where the virtual photon is at rest
and that we denote the hyperons' momentum by $p_z$, cf.\ \eqref{eq:pzdef}. 

In subsection \ref{sec:opp-par3/2} we discuss the case of opposite parity of initial spin-3/2 and final
spin-1/2 hyperon. In this case, the orbital angular momentum $l$ must be even. In total we find three partial waves:
$l=0$ and $s=3/2$; $l=2$ and $s=1/2$; $l=2$ and $s=3/2$. Here $s$ denotes the total spin built from the spin 1 of the photon
and the spin 1/2 of the final hyperon. Since the amplitude for orbital angular momentum $l$ scales with $p_z^l$, one amplitude
(the \textit{s}-wave) becomes dominant at the end of the phase space. This must find its expression in a relation between the
helicity amplitudes. Indeed, this is what the kinematical constraint \eqref{eq:helconstm3/2} means physically.

To complete this discussion, we note that due to crossing symmetry the same amplitude that describes
$Y^* \to Y \, e^+ e^-$ can also be used to describe $e^+ e^- \to Y \bar Y^*$. A non-relativistic situation emerges at the
production threshold $q^2 \approx (m_{Y^*}+m_Y)^2$. Since the antifermion $\bar Y^*$ has opposite parity to $Y^*$, the allowed
orbital angular momentum $l$ must now be odd. The three possible partial waves are $l=1$ and $s=1$; $l=1$ and $s=2$;
$l=3$ and $s=2$,
where now $s$ denotes the total spin built from the spin 3/2 of the antihyperon and the spin 1/2 of the hyperon. At threshold, 
the two \textit{p}-waves dominate over the \textit{f}-wave. Thus one of the three helicity amplitudes must be related to the other two. This
is the physics content of the kinematical constraint \eqref{eq:helconstm3/2-high}.

Analogous considerations can be used for the other spin-parity combinations that we discuss in this paper. For every case, one
can understand where and how many kinematical constraints emerge for the helicity amplitudes.

\section{General structure of the squared matrix element for the four-body decay}
\label{app:4body}

We consider the decay $Y^* \to Y \, e^+ e^-\to N \pi \, e^+ e^-$. The spin summed/averaged squared matrix element 
$\langle \vert {\cal M}_4 \vert^2 \rangle$ can be expressed in terms of the
four four-vectors $p_Y$, $p_N$, $q=p_{e^-}+p_{e^+}$ and $k_e=p_{e^-}-p_{e^+}$.
The Lorentz invariant combinations that cannot be
expressed solely by masses are $q^2$, $k_e \cdot p_Y$, $q \cdot p_N$, $k_e \cdot p_N$ and
$\epsilon_{\mu\nu\alpha\beta} \, k_e^\mu \, p_Y^\nu \, p_N^\alpha \, q^\beta$. 

In the following we will show that the general structure is
\begin{eqnarray}
  \langle \vert {\cal M}_4 \vert^2 \rangle &=& J_1(q^2) + J_2(q^2) \, (k_e \cdot p_Y)^2 \nonumber \\
  && {} + J_3(q^2) \, k_e \cdot p_Y \;
  \epsilon_{\mu\nu\alpha\beta} \, k_e^\mu \, p_Y^\nu \, p_N^\alpha \, q^\beta  \,.
  \label{eq:genM4}  
\end{eqnarray}
In particular, we will show that
$\langle \vert {\cal M}_4 \vert^2 \rangle$ does {\em not} depend on $q \cdot p_N$ or $k_e \cdot p_N$. 

An evaluation of the lepton trace shows that
$k_e$ can only appear pairwise. It is also easy to see that $p_N$ can appear at most linearly. This defines already the
structure with the Levi-Civita tensor. No terms $\sim q \cdot p_N, k_e \cdot p_N$ could show up there, because there is already
one $p_N$ contracted with the Levi-Civita tensor. It remains to be shown that any term without a Levi-Civita tensor does not 
depend on $q \cdot p_N$ or $k_e \cdot p_N$. 

The Levi-Civita structure emerges from an odd \newline number of $\gamma_5$ matrices while any other structure 
stems from terms with an even number of $\gamma_5$ matrices. Consequently, the weak coupling constants $A$ and $B$ from
\eqref{eq:Lambdadecay} appear each linearly in $J_3$, while the contributions to any other structure are $\sim \vert A \vert^2$
or $\sim \vert B \vert^2$. Thus one has to deal there either with the $A$-type interaction or with
the $B$-type interaction, but not with both at the same time.
Both interactions are formally parity invariant if one considers the nucleon as having opposite or
the same parity as the $Y$. Thus one can use parity arguments to show that any term that does not contain the
Levi-Civita tensor will be independent of $q \cdot p_N$ and $k_e \cdot p_N$.

Now we consider the rest frame of the $Y$ and a spin quantization axis along the motion of the nucleon. In this frame,
the spin of the $Y$ must be identical to the spin of the nucleon. Let ${\cal M}_w(s_N,s_Y)$ be the decay matrix
element caused by the $A$- {\em or} $B$-interaction (exclusive ``or''). Here $s_{N/Y}$ is the spin of the
nucleon or $Y$, respectively. We find
\begin{equation}
    \begin{split}
    \qquad \qquad\quad {\cal M}_w(s_N,s_Y) \sim{}& \delta_{s_N,s_Y} \,,  \\
  \vert {\cal M}_w(+1/2,+1/2) \vert ={}& \vert {\cal M}_w(-1/2,-1/2) \vert \,. 
    \end{split}
  \label{eq:Mweak-AorB}
\end{equation}

Let ${\cal M}_s(\ldots,s_Y)$ denote the decay matrix element for the decay $Y^* \to Y \, e^+ e^-$. The dots
denote the spins of the other states. We do not need to specify them further. In total we find
\begin{eqnarray}
  \label{eq:longcalcM4}
  &&  \langle \vert {\cal M}_4 \vert^2 \rangle_{A \; {\rm or} \, B}  
  \\
  &\sim& \sum\limits_{\ldots} \sum\limits_{s_N, s_Y,s'_Y} {\cal M}_w(s_N,s_Y) \, {\cal M}_s(\ldots,s_Y)  \nonumber \\
  && \phantom{mmmmmm} \times {\cal M}_s^*(\ldots,s'_Y) \, {\cal M}^*_w(s_N,s'_Y)
         \nonumber \\[1em]
  &=& \sum\limits_{\ldots} \sum\limits_{s_N} {\cal M}_w(s_N,s_N) \, {\cal M}^*_w(s_N,s_N) \nonumber \\
  && \phantom{mmmm} \times {\cal M}_s(\ldots,s_N) \, {\cal M}_s^*(\ldots,s_N)
      \nonumber \\[1em]
  &=& \sum\limits_{\ldots} \sum\limits_{s_N} \vert {\cal M}_w(s_N,s_N) \vert^2 \, \vert {\cal M}_s(\ldots,s_N) \vert^2
      \nonumber \\
  &=& \sum\limits_{\ldots} \left(\vert {\cal M}_w(+1/2,+1/2) \vert^2 \, \vert {\cal M}_s(\ldots,+1/2) \vert^2 
  \right. \nonumber \\  && \phantom{mm} \left.
      + \vert {\cal M}_w(-1/2,-1/2) \vert^2 \, \vert {\cal M}_s(\ldots,-1/2) \vert^2 \right)
      \nonumber \\[1em]
  &=& \vert {\cal M}_w(+1/2,+1/2) \vert^2 \nonumber \\
  && \times \sum\limits_{\ldots} \left(
      \vert {\cal M}_s(\ldots,+1/2) \vert^2 + \vert {\cal M}_s(\ldots,-1/2) \vert^2 \right)
      \nonumber \\
  &=& \frac12 \left( \sum\limits_{s_N,s_Y} \vert {\cal M}_w(s_N,s_Y) \vert^2 \right)
      \left( \sum\limits_{s'_Y,\ldots} \vert {\cal M}_s(\ldots,s'_Y) \vert^2 \right) \,.  \nonumber 
\end{eqnarray}
Now we have reached a product of two spin-summed squares of matrix elements. Both are Lorentz invariant and can be
evaluated in any frame. The first one does not depend on $q$ or $k_e$. The second does not depend on $p_N$. Therefore,
the products $q \cdot p_N$ and $k_e \cdot p_N$ do not appear. 

We note again that this line of reasoning only works if one considers solely the $A$-type interaction or solely the
$B$-type interaction. It does not work for interference terms $\sim A B^*, A^* B$. But such interference terms are accompanied
by an odd number of $\gamma_5$ matrices and therefore give rise to a Levi-Civita tensor.

\bibliography{lit}{}
\bibliographystyle{utphysmod}
\end{document}